\documentclass{ws-ijmpa}
\usepackage[super,compress]{cite}
\usepackage{graphicx}
\usepackage[breaklinks]{hyperref}

\newcommand{\bbar}{\ensuremath{\bar{\rm b}}}
\newcommand{\cbar}{\ensuremath{\bar{\rm c}}}
\newcommand{\sbar}{\ensuremath{\bar{\rm s}}}
\newcommand{\pbar}{\ensuremath{\bar{\rm p}}}
\newcommand{\tbar}{\ensuremath{\bar{\rm t}}}
\newcommand{\qbar}{\ensuremath{\bar{\rm q}}}

\begin{document}
\markboth{M.~Voutilainen}
{HEAVY QUARK JETS AT THE LHC}

%
\catchline{}{}{}{}{}
%

\title{
HEAVY QUARK JETS AT THE LHC
}

\author{
MIKKO VOUTILAINEN
}

\address{
Helsinki Institute of Physics\\
P.O. Box 64, University of Helsinki
Finland
\\
mikko.voutilainen@cern.ch}



\maketitle

\begin{history}
\received{Day Month Year}
\revised{Day Month Year}
\end{history}

\begin{abstract}
We summarize measurements of b and c jet production at the LHC, which are an important signature and background for decays of massive particles such as H$\rightarrow$b$\bar{\rm b}$. These include measurements of the inclusive and dijet production of heavy quark jets, b and c jets produced in association with vector bosons Z and W, and decays of boosted Z bosons into pairs of b$\bar{\rm b}$. The current status of b tagging and b jet energy scale is also reviewed.
These measurements test perturbative QCD in the four and five-flavor number schemes, and provide insight into the relative importance of heavy flavor production through flavor creation, flavor excitation and gluon splitting channels. The W+c measurement provides additionally a powerful way to probe the strange quark and antiquark sea in the proton.
The recent studies looking separately at production of one and two b jets find generally good agreement with theory predictions for two b-jet production, while some discrepancies are observed for singly produced b jets, particularly at large b-jet $p_T$, where gluon splitting becomes dominant.
\end{abstract}

\ccode{PACS numbers:}


\section{Introduction}

\begin{figure}[htbp!]
\centering
\includegraphics[width=0.32\textwidth]{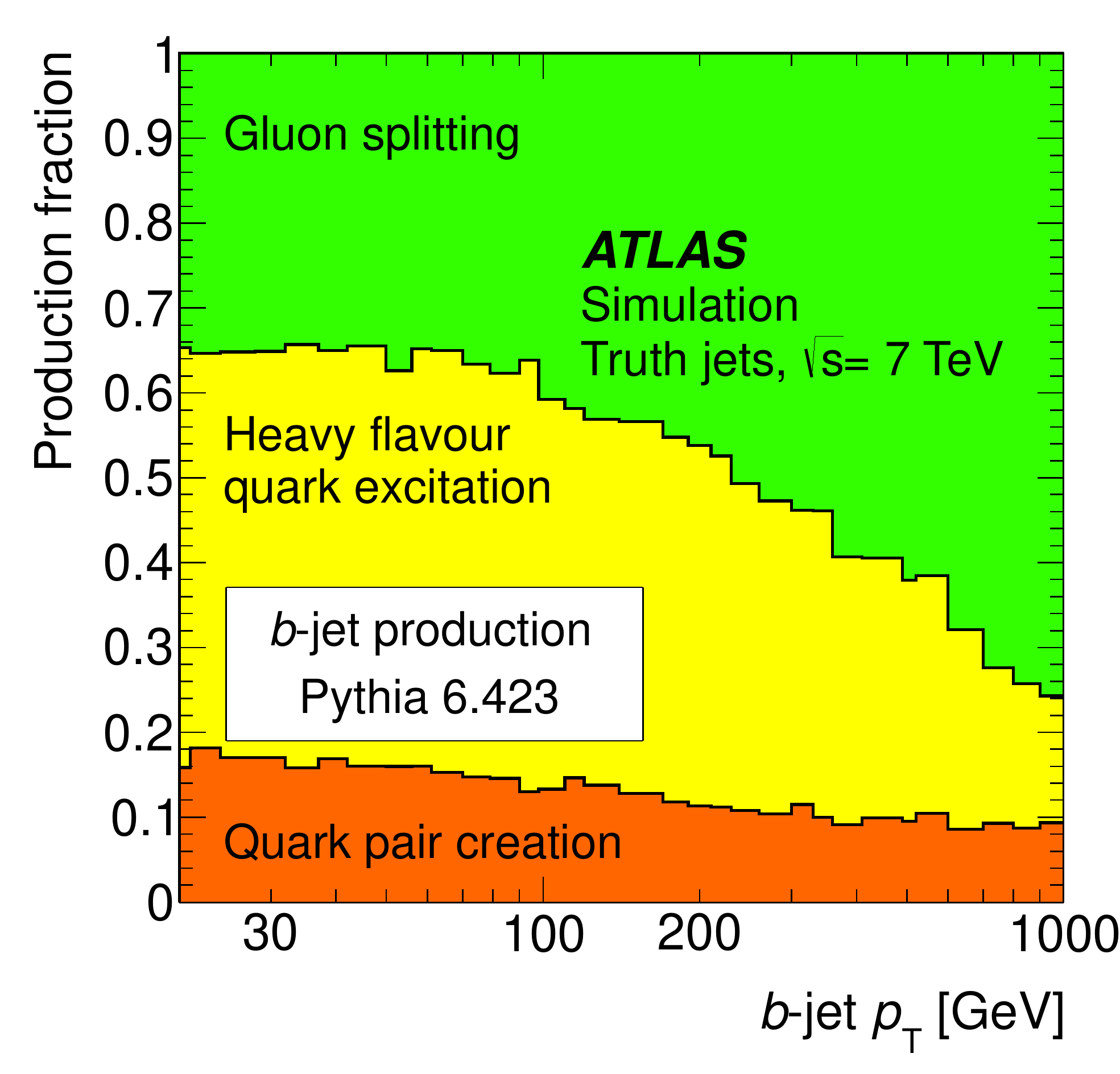}
\includegraphics[width=0.32\textwidth]{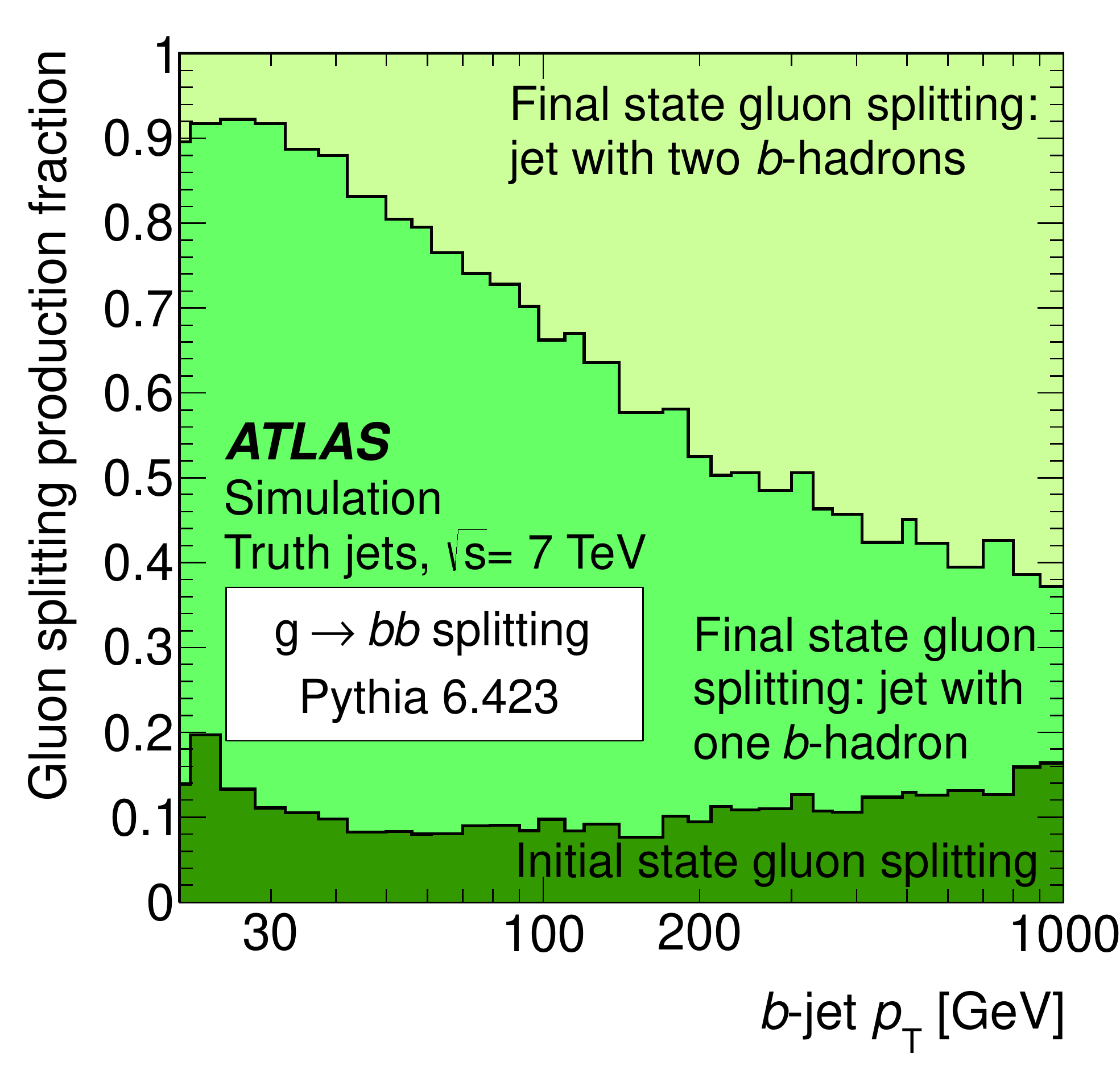}
\includegraphics[width=0.32\textwidth]{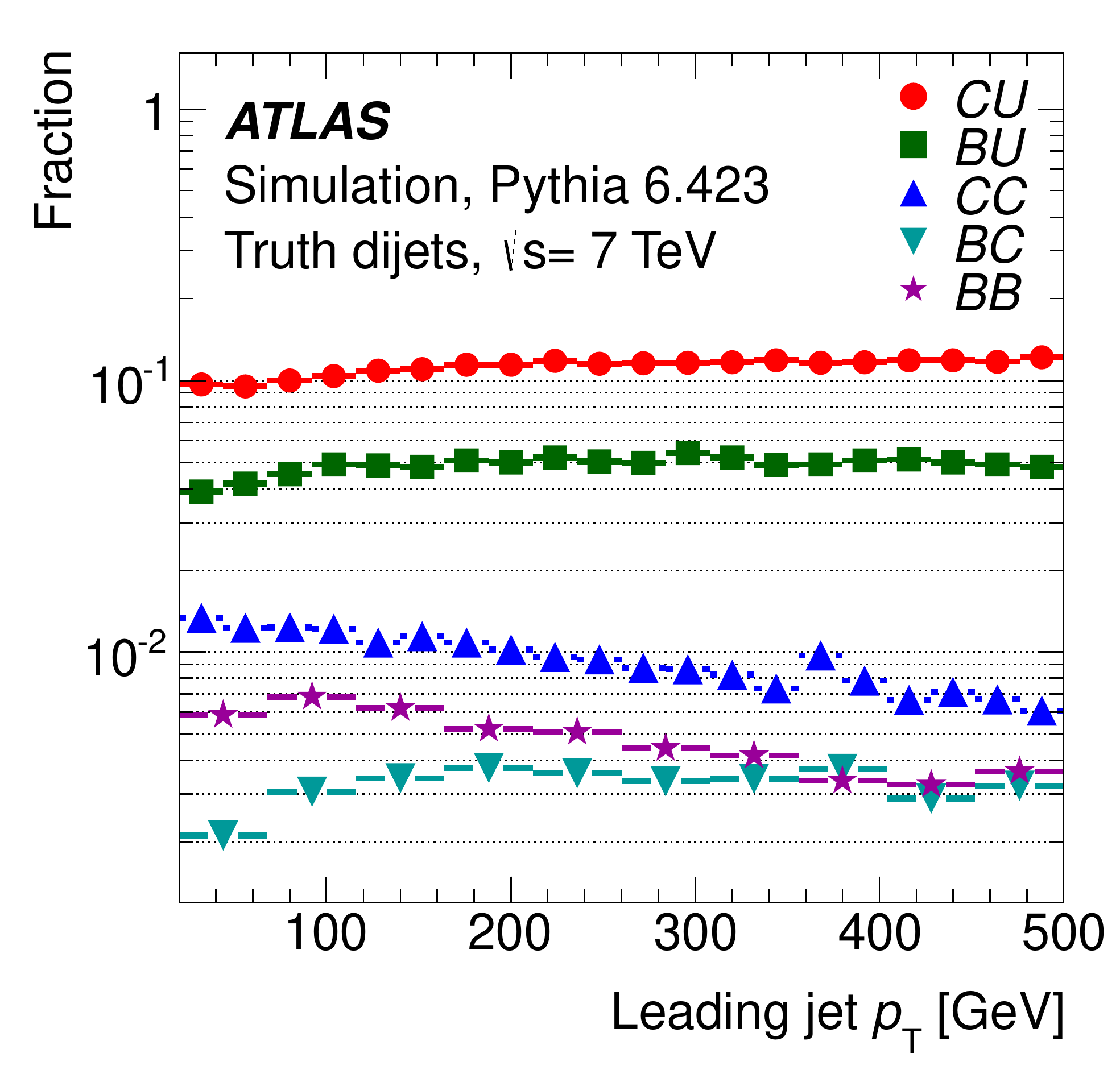}
\caption{\label{fig:bbfrac} (Left) Fraction of b jets produced through different channels, and (middle) sub-division of gluon splitting production fraction by number of b hadrons in a jet. 
(Right) Predicted fraction of jet flavor pairs in dijet events as a function of leading jet $p_T$.
\cite{Aad:2012ma}
}
\end{figure}

Measurements of heavy flavor jet production with and without vector bosons offer insights into the proton parton distribution functions (PDFs), and an important testing ground of perturbative QCD (pQCD) predictions in the presence of massive quarks. 
The b and c jets are produced in substantial amounts ($\sim$4\% b, $\sim$10\% c) in strong interactions \cite{Chatrchyan:2012dk,ATLAS:2011ac,Aad:2012ma} through processes often referred to as flavor creation (FC), flavor excitation (FEX) and gluon splitting (GS), which form an important background to electroweak production of heavy flavors. The relative proportions of these processes are shown in Fig.~\ref{fig:bbfrac}.
Flavor composition of dijet events \cite{Aad:2012ma} allows to discriminate between the various b-jet production modes. This is particularly useful for the dominant GS channel that requires next-to-leading order (NLO) calculations for precise modeling.

The c-jets produced in association with a W boson \cite{Chatrchyan:2013uja,Aad:2014xca} are a powerful way to probe the strange quark and antiquark sea of the proton, while W+b \cite{Aad:2011kp,Aad:2013vka,Chatrchyan:2013uza} and Z+b \cite{Aad:2011jn,Aad:2014dvb,Chatrchyan:2012vr,Chatrchyan:2013zja,Chatrchyan:2014dha} test the production of heavy quarks in pQCD. The b quarks are included in proton parton distribution functions (PDFs) only in the five-flavor number scheme (5FNS) so W+b and Z+b provide a way to compare 5FNS to the four-flavor number scheme (4FNS), in which b jets are primarily produced in pairs through gluon splitting.

The leading NLO pQCD predictions on the market are MCFM \cite{Campbell:2010ff} and aMC@NLO \cite{Alwall:2014hca}, which implements both 4FNS and 5FNS, and leading multileg leading-order (LO) Monte Carlo (MC) generators are MadGraph \cite{Alwall:2011uj}, Sherpa \cite{Gleisberg:2008ta} and AlpGen \cite{Mangano:2002ea}, each interfaced with Pythia \cite{Sjostrand:2006za} or Herwig \cite{Bahr:2008pv}. The various theoretical approaches and their benefits and drawbacks are reviewed {\em e.g.} by Maltoni {\em et al}~\cite{Maltoni:2012pa}.

The b jets, and to a lesser extent c jets, are an important signature of decays of massive particles such as t$\rightarrow$Wb, H$\rightarrow$b\bbar, Z$\rightarrow$b\bbar\ and W$\rightarrow$c\sbar. 
Of particular importance to the ongoing LHC physics program is the identification of a Higgs boson decaying to two b jets. Because of the large strong interaction background of heavy flavor jets, this is best achieved by looking at the associated production of a Higgs boson with a vector boson (W, Z), or by tagging a highly boosted Higgs boson produced in association with two forward quark jets in the vector boson fusion (VBF) channel. Important benchmarks for these channels are multiple b jets produced in association with W \cite{Chatrchyan:2013uza,Aad:2013vka} and Z bosons\cite{Aad:2011jn,Aad:2014dvb,Chatrchyan:2012vr,Chatrchyan:2013zja,Chatrchyan:2014dha}, and b jets produced from boosted Z$\rightarrow$b\bbar\ decays \cite{Aad:2014bla}.

The b jets produced in cascade decays of {\em e.g.} supersymmetric particles are an important indicator of the presence of heavy particles with suitable properties, and for this reason b tagging is often used to enhance new physics searches.

From an analysis perspective, the heavy flavor (b, c) jets are unique among jet flavors in that they are experimentally identifiable with high efficiency and purity. The b jets in particular can be tagged with efficiencies higher than 70\% and with light flavor mistag rates of the order of one percent \cite{Chatrchyan:2012jua,CMS-PAS-BTV-13-001,ATLAS-CONF-2010-042,ATLAS-CONF-2010-091,ATLAS-CONF-2011-089, ATLAS-CONF-2011-102,ATLAS-CONF-2012-043,ATLAS-CONF-2012-097,ATLAS-CONF-2014-004}. Typical heavy flavor tagging methods rely on the identification of secondary vertices, track impact parameters and soft leptons, or combinations thereof.

The fragmentation properties of b and c jets are in between those of light quarks and gluons, owing to the high mass of the b and c hadrons. The neutrinos produced in the semileptonic decays of the heavy hadrons are a peculiarity that presents particular experimental challenges. Even excluding neutrinos, the energy scale of b jets \cite{Chatrchyan:2011ds,CMS-DP-2012-006,Aad:2014bia} is among the leading experimental systematics for many precision studies involving b jets.

In this article we will first review the experimental aspects of heavy flavor measurements, b tagging (Sec.~\ref{sec:btag}) and b-jet energy scale (Sec.~\ref{sec:bjes}), then move on to measurements of inclusive b-jet and dijet b cross sections (Sec.~\ref{sec:bjet}).
The W+c (Sec.~\ref{sec:Wc}), V+b (Sec.~\ref{sec:Vb}) processes are reserved a section each, 
before reviewing measurements of Z boson decay in boosted topologies (Sec.~\ref{sec:Ztobb}). The conclusions and outlook are given in Sec.~\ref{sec:summary}.

\section{B tagging}
\label{sec:btag}

The heavy flavor jets (b, c) are typically defined as containing a heavy hadron or a b/c quark from parton shower. These definitions include jets produced from gluon splitting (g$\rightarrow$b\bbar, g$\rightarrow$c\cbar), which constitute an irreducible background for tagging algorithms.
The b tagging relies on properties of the b hadrons: large mass, long lifetime and daughter particles with hard spectra. Tagging is also possible using semileptonic decays and fully reconstructed heavy hadron decays. The c tagging is typically not done directly, but as a side product of b tagging.

\begin{figure}[htbp!]
\centering
\includegraphics[width=0.49\textwidth]{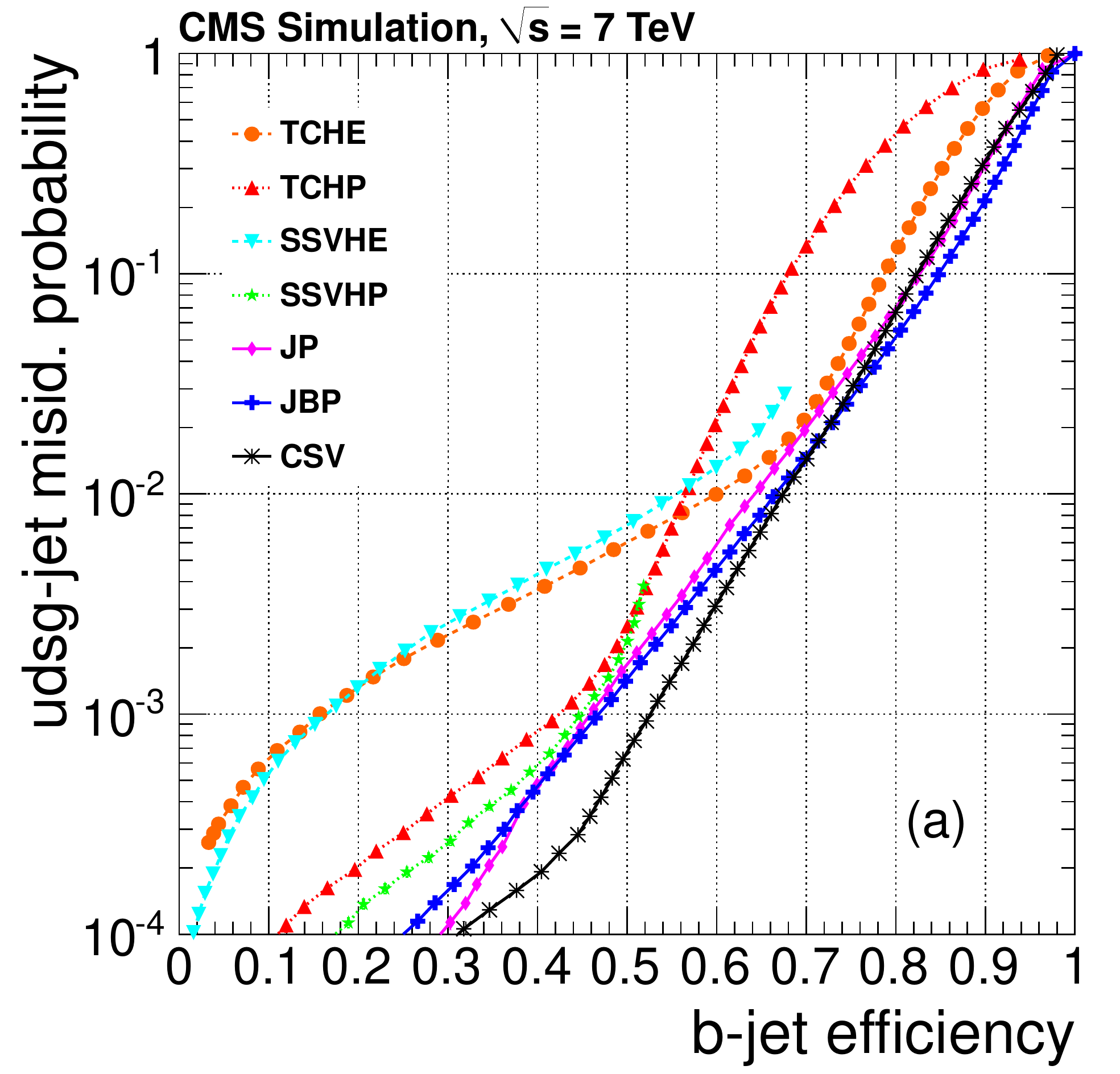}
\includegraphics[width=0.49\textwidth]{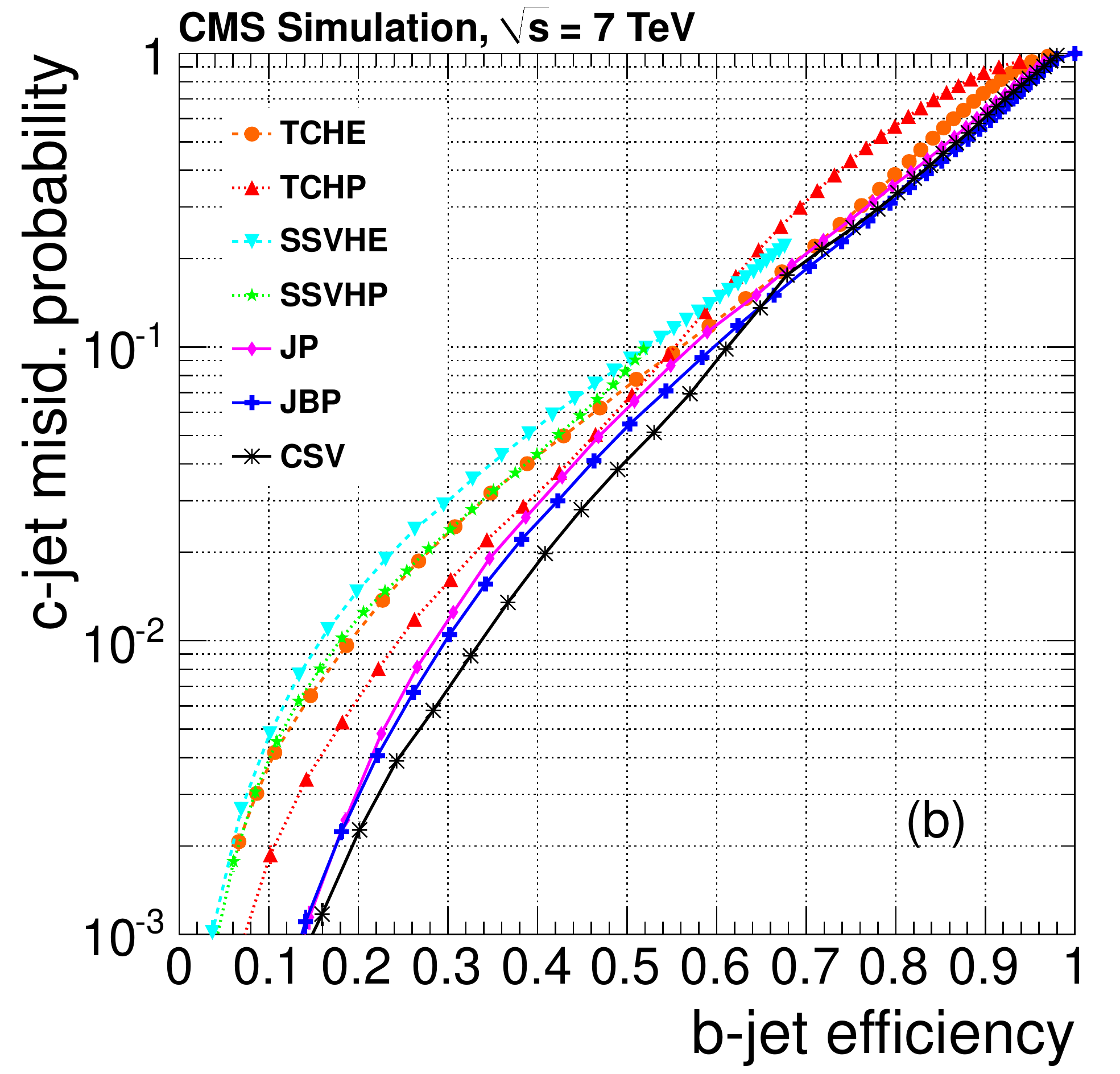}
\caption{\label{fig:btag}
(Left) Light-parton jet misidentification probability versus b-jet efficiency.
(Right) Charm-jet misidentification probability versus b-jet efficiency.
\cite{Chatrchyan:2012jua}
}
\end{figure}

CMS and ATLAS have implemented multiple b-taggers using track impact parameters (TCH; JetProb, IP3D), properties of decay vertices (SSV; SV0, SV1), presence or absence of leptons, and combinations thereof (JP, JBP, CSV; JetFitter, MV1). Typical b-tagging efficiencies with respect to light-parton (u, d, s, g) misidentification probability are 85\% versus 10\% for high efficiency (HE) working points, 70\% versus 1\% for medium working points, and 50\% versus 0.1\% for high purity (HP) working points. The performance of various algorithms at CMS is summarized in Fig.~\ref{fig:btag} \cite{Chatrchyan:2012jua}. ATLAS' performance is similar.

The best results are obtained with multivariate methods, but many analyses reported in this review rely on the medium or high efficiency working point of the simple and robust secondary vertex lifetime tagging to control systematic uncertainties. In addition, one recent analysis \cite{Aad:2012ma} uses flavor templates without explicit b tagging, and demonstrates ability to separate prompt b quark production from the gluon-splitting contribution.

Physics analyses correct for data/simulation differences in b-tagging efficiency and light-parton misidentification probability using scale factors $SF_{\rm b}$ and $SF_{\rm light}$, respectively. These scale factors and their uncertainties are determined from data using multiple complementary methods\cite{Chatrchyan:2012jua,ATLAS-CONF-2012-043}. The charm efficiency is assumed the same as b, with larger uncertainty. Only CMS has published results on b-tagging performance in data \cite{Chatrchyan:2012jua,CMS-PAS-BTV-13-001}, while ATLAS has released a series of conference notes \cite{ATLAS-CONF-2010-042,ATLAS-CONF-2010-091,ATLAS-CONF-2011-089, ATLAS-CONF-2011-102,ATLAS-CONF-2012-043,ATLAS-CONF-2012-097,ATLAS-CONF-2014-004}.

\begin{figure}[htbp!]
\centering
\includegraphics*[width=0.43\textwidth, viewport=0 20 540 280]{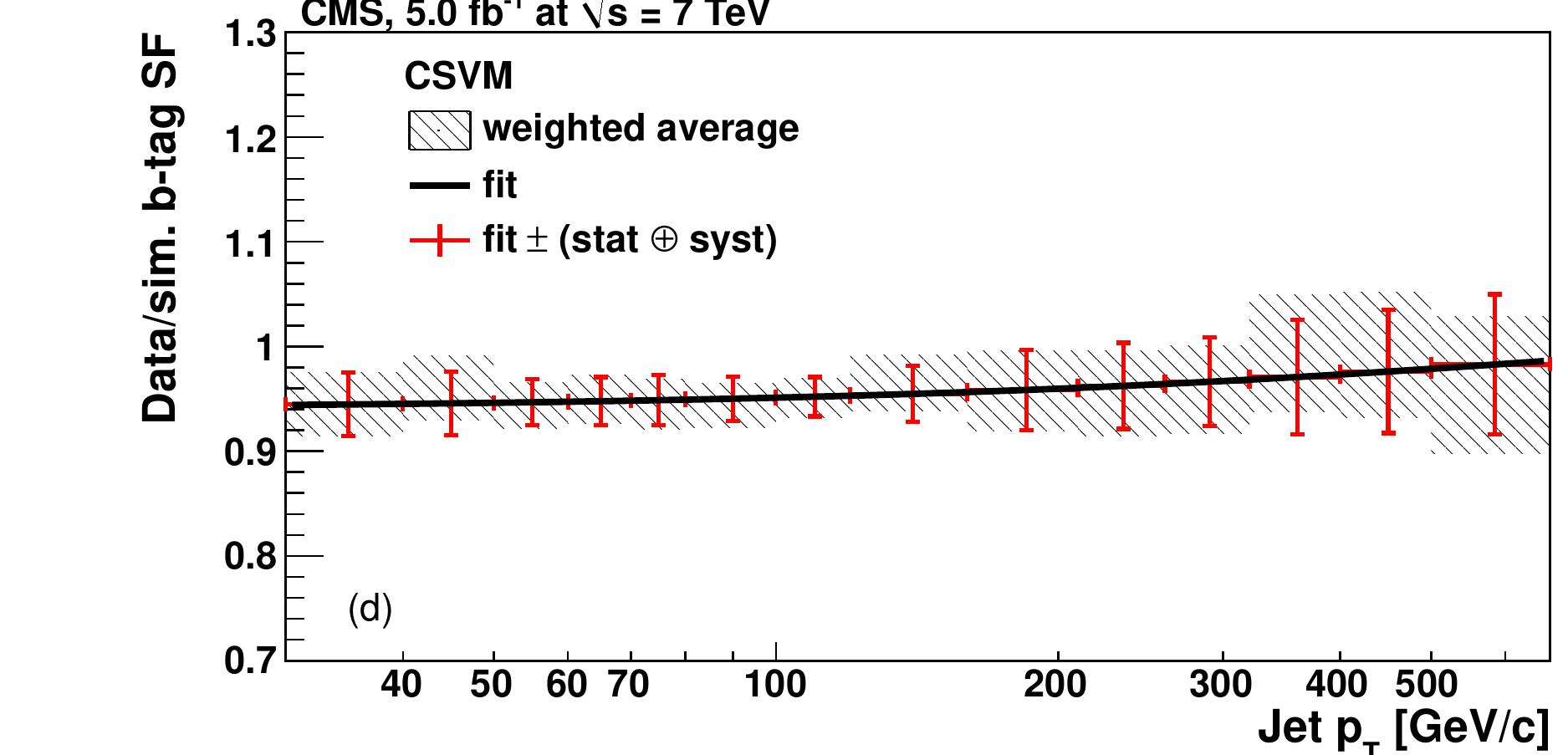}
\includegraphics*[width=0.55\textwidth, viewport=0 0 520 190]{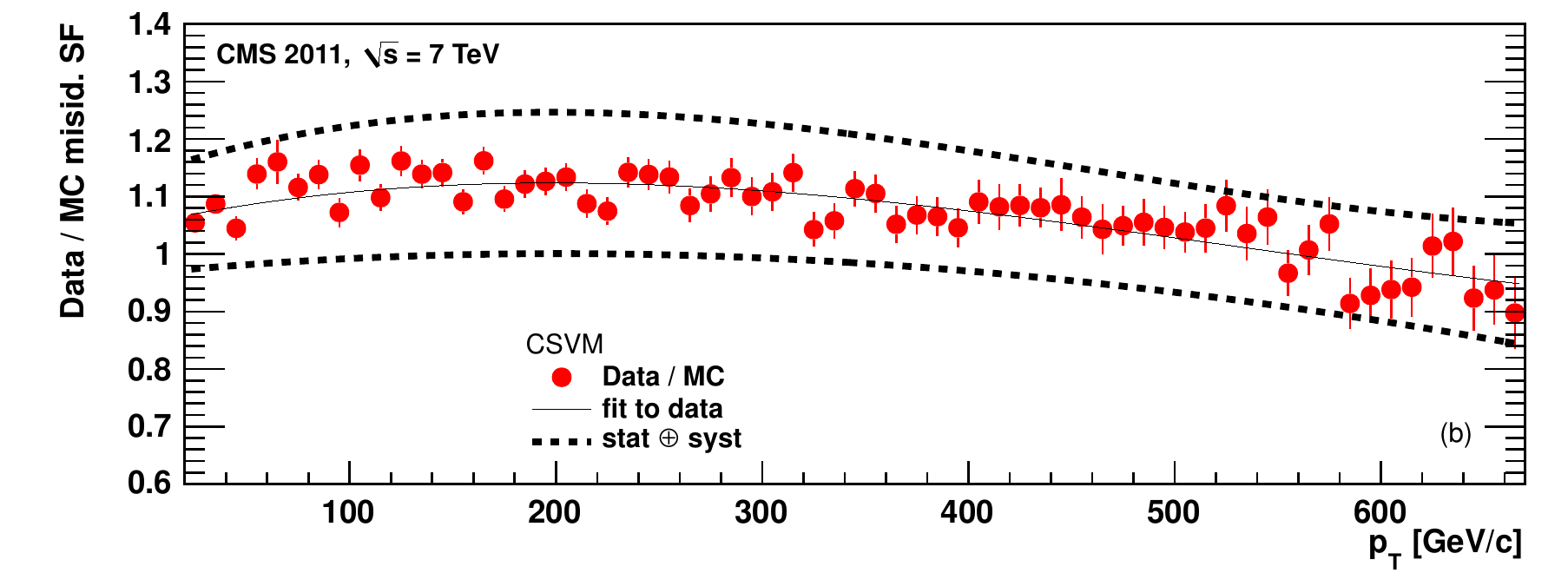}
\caption{\label{fig:btagsf}
(Left) Data/simulation scale factor for b-tagging efficiency, and
(right) data/simulation scale factor for light-parton misidentification probability with CSVM tagger.
\cite{Chatrchyan:2012jua}
}
\end{figure}

Both experiments find general agreement with simulation, with $SF_b$ varying between 0.9--1.0 and with uncertainty of 2--4\% for the latest measurements. The light-parton misidentification probability scale factor is at most 25\% different from unity with 8--17\% uncertainty. The efficiencies and scale factors depend on $\eta$ and $p_T$ of the jets. The CSV medium working point (CSVM) performance in data at CMS is shown in Fig.~\ref{fig:btagsf} as an example.

\section{B-jet energy scale}
\label{sec:bjes}

\begin{figure}[htbp!]
\centering
\includegraphics[width=0.49\textwidth]{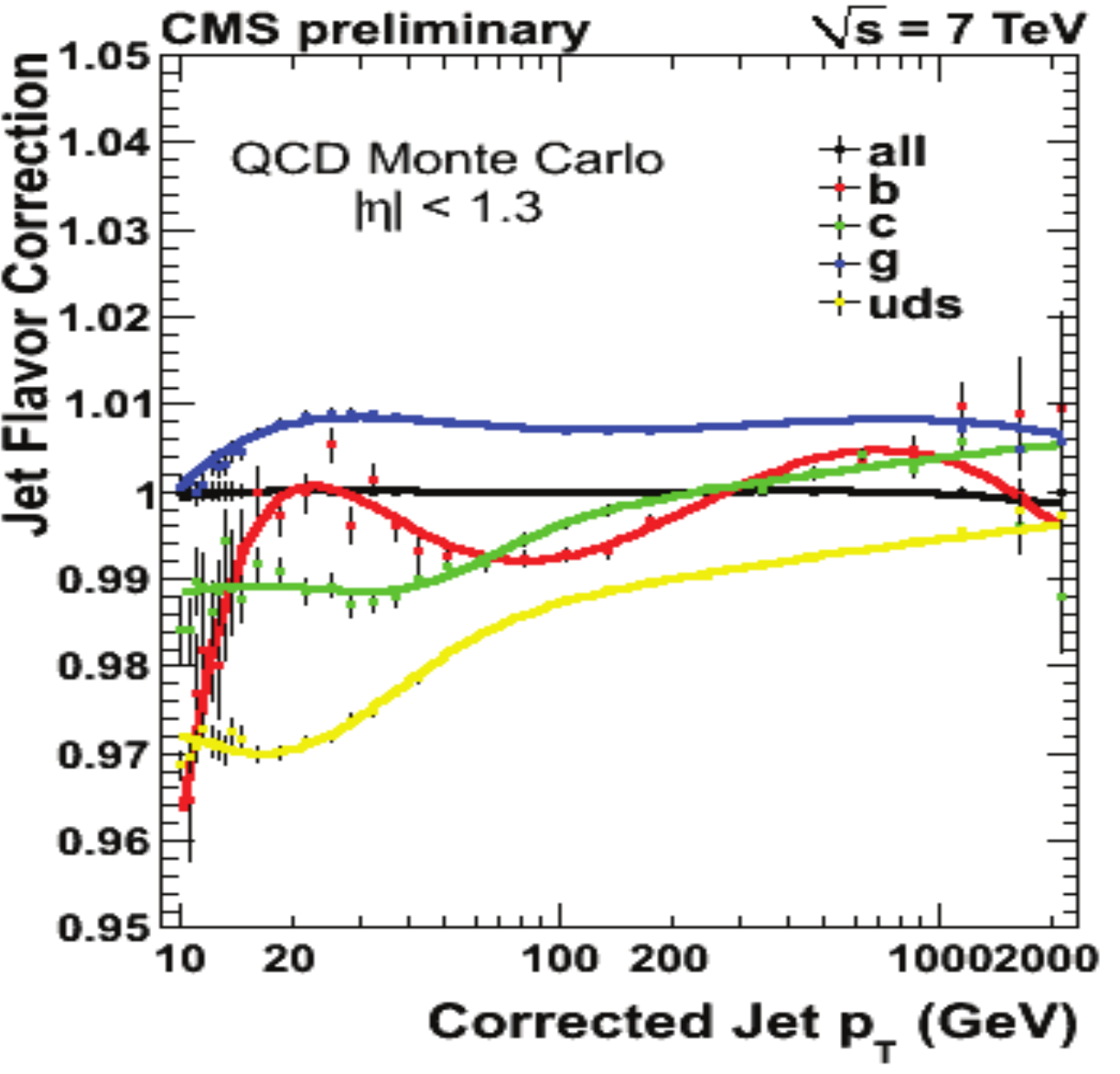}
\includegraphics[width=0.49\textwidth]{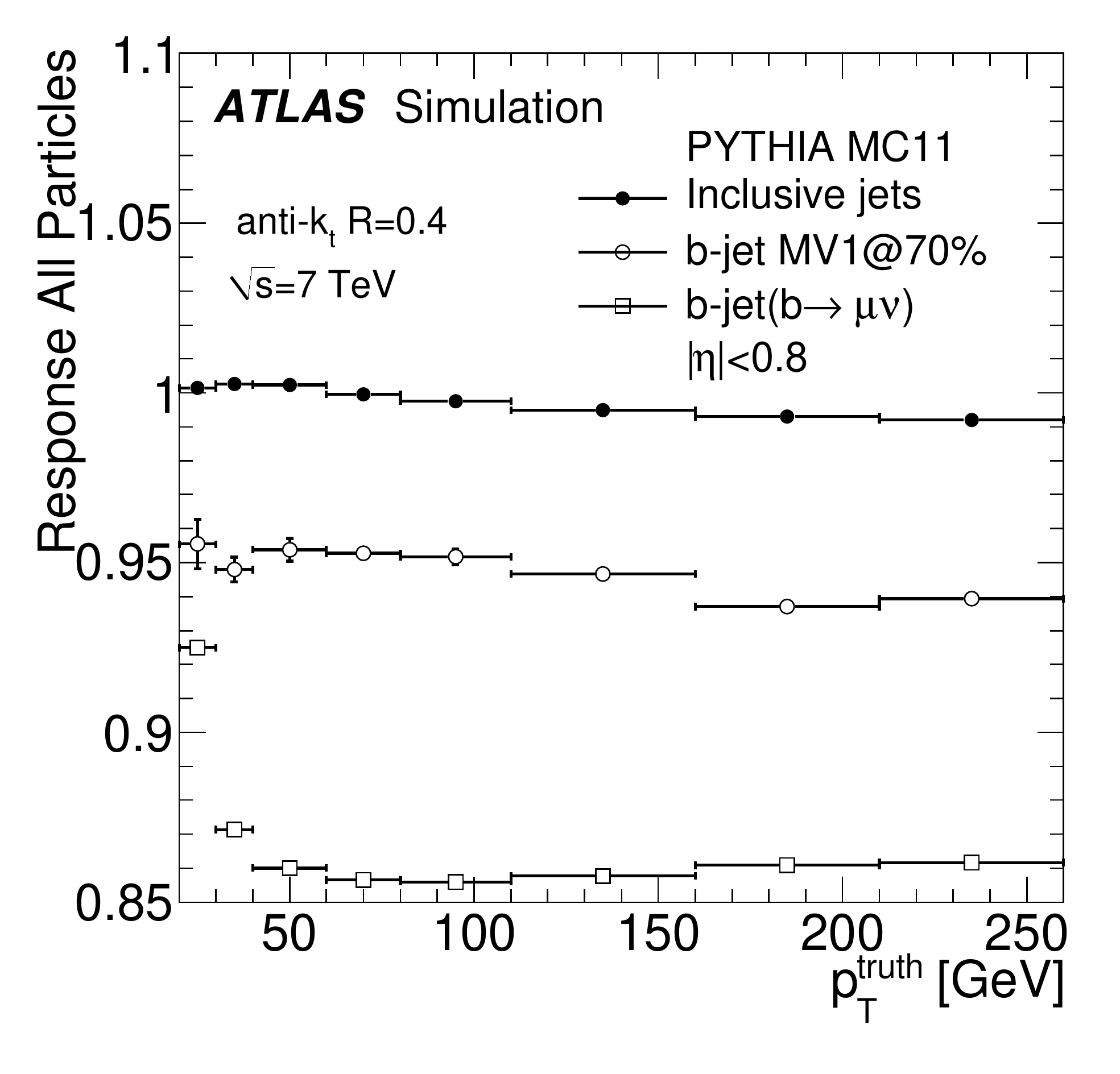}
\caption{\label{fig:bjes}
(Left) The b-jet energy correction relative to a QCD dijet mixture of jets, compared to light quarks (uds), charm (c) and gluons (g). The reference particle jets do not contain neutrinos.
\cite{CMS-DP-2012-006}
(Right) The b-jet response to all particles including neutrinos, and for a subsample of b jets tagged with semileptonic decays.
\cite{Aad:2014bia}
}
\end{figure}

The b-jet energy scale (b-JES) is one of the leading systematics for many measurements of b jets, including those in this review. The multiple hadrons produced in b hadron decays result in softer fragmentation than that of light quark jets, but not quite as soft as that of gluon jets, even for b jets produced through gluon splitting. Excluding neutrinos produced in semileptonic decays, the b-jet response is therefore expected to be between uds and gluon jets. It is also coincidentally quite close to the inclusive jet energy scale (JES), as shown in Fig.~\ref{fig:bjes}(left). The b-JES is discussed in the main JES publications of CMS \cite{Chatrchyan:2011ds,CMS-DP-2012-006} and ATLAS \cite{Aad:2014bia}, with a dedicated study performed with Z+b events at 8~TeV by CMS \cite{CMS-PAS-JME-13-001}.

The neutrinos produced in semileptonic decays of b hadrons are not detected, which leads to an average energy loss of about 5\% when comparing to the energy of the b quarks or particle jets clustered with neutrinos, as shown in Fig.~\ref{fig:bjes}(right). The semileptonic decays account for about 20\% of direct b-hadron decays, and another 30\% of cascade decays of c hadrons.
For b jets tagged with soft muons the energy lost in neutrinos corresponds up to 15\% of the original parton energy. The energy carried by neutrinos is found to be well modeled in the simulation, within an uncertainty of about 2\% \cite{Aad:2014bia}.

The uncertainty in the b-jet response relative to inclusive jets is dominated by the b-jet fragmentation, estimated by comparing various MC generators.
CMS assigns a flavor uncertainty based on differences in the jet response relative to the inclusive jet sample, estimated by comparing Pythia 6 and Herwig++. This jet flavor uncertainty is based on the maximal uncertainty envelope of all flavors, which is dominated by gluons, and is estimated to be 0.5--1.5\%. ATLAS has performed a similar study, and finds the relative uncertainty of b jets versus inclusive jets to be in the same range.

The most accurate data-based constraints on the b-JES relative to inclusive JES to date have been obtained by comparing Z+b-jet events to inclusive Z+jet events \cite{CMS-PAS-JME-13-001},
as discussed in the review article on jet energy corrections at the LHC. 
This study finds a residual b-JES correction of $C_{\rm corr} = 0.998\pm 0.004(\rm stat.)\pm 0.004(\rm syst.)$ relative to Pythia 6 and $C_{\rm corr} = 1.005\pm 0.006(\rm stat.)\pm 0.004(\rm syst.)$ relative to Herwig++, consistent with no correction in both cases, and compatible with the 0.5--1.5\% uncertainty on b-JES relative to inclusive JES obtained from fragmentation studies in simulation.

Another potentially useful channel for constraining b-JES for higher $p_T$ b jets in the future is the boosted decay of Z$\rightarrow$b\bbar\ measured by ATLAS, discussed in Sec.~\ref{sec:Ztobb}. This study finds $\delta M_Z = -1.5\pm 0.7(\rm stat.)^{+3.4}_{-2.5}(\rm syst.)$~GeV, which is also consistent with no correction.

\section{Inclusive and dijet b production}
\label{sec:bjet}

The production of b quarks at the LHC provides an important test of pQCD. The early measurements from Sp\pbar{}S, Tevatron and HERA (see {\em e.g.} \cite{Lourenco:2006vw} and references therein) found discrepancies between data and predictions which led to substantial amount of work on the theory side. The subsequent improvements led to reasonable agreement, but still with sizable theoretical uncertainties. Therefore, there is great interest to test the theoretical predictions at the higher energy range of the LHC.

The total b-hadron cross section has been measured by LHCb using semi-inclusive decays in the forward rapidity regions \cite{Aaij:2010gn} and by CMS using inclusive b$\rightarrow{\rm\mu}$X decays \cite{Khachatryan:2011hf}. CMS has also measured production of fully reconstructed B$^+$ \cite{Khachatryan:2011mk}, B$^{0}$ \cite{Chatrchyan:2011pw}, and B$_{\rm s}$ \cite{Chatrchyan:2011vh} mesons, as well as the angular correlations between b and \bbar\ mesons \cite{Khachatryan:2011wq}.
These have found the measured cross sections to lie between MC@NLO and Pythia predictions.

The b (about 5~GeV) and c (about 1~GeV) quark masses are above typical QCD scales ($\Lambda_{\rm QCD}\approx 200$~MeV) so their production is not affected by low energy hadronization effects and is directly described by pQCD. The b-hadron identification by lifetime is also not sensitive to fragmentation effects. Therefore measurement of b jets is a direct measurement of the b-quark production with negligible systematic uncertainty from fragmentation.
Both ATLAS and CMS collaborations have measured the inclusive cross sections of b jets \cite{ATLAS:2011ac,Chatrchyan:2012dk}, and ATLAS has also measured dijet cross sections of b jets \cite{ATLAS:2011ac}, including their flavor compositions \cite{Aad:2012ma}. The latter analysis uses a unique fit approach with no explicit b tagging.

The measurement of inclusive b jets avoids large logarithmic corrections to the theory predictions due to hard collinear gluons, while the measurements of the di-b-jet cross section have reduced contributions from gluon splitting. The exclusive dijet cross sections for b and c jets in association with lighter flavors allow separating the different heavy flavor quark creation processes, providing more detailed information about the different QCD processes involving heavy quarks.

\begin{figure}[htbp!]
\centering
\includegraphics[width=0.46\textwidth, angle=90]{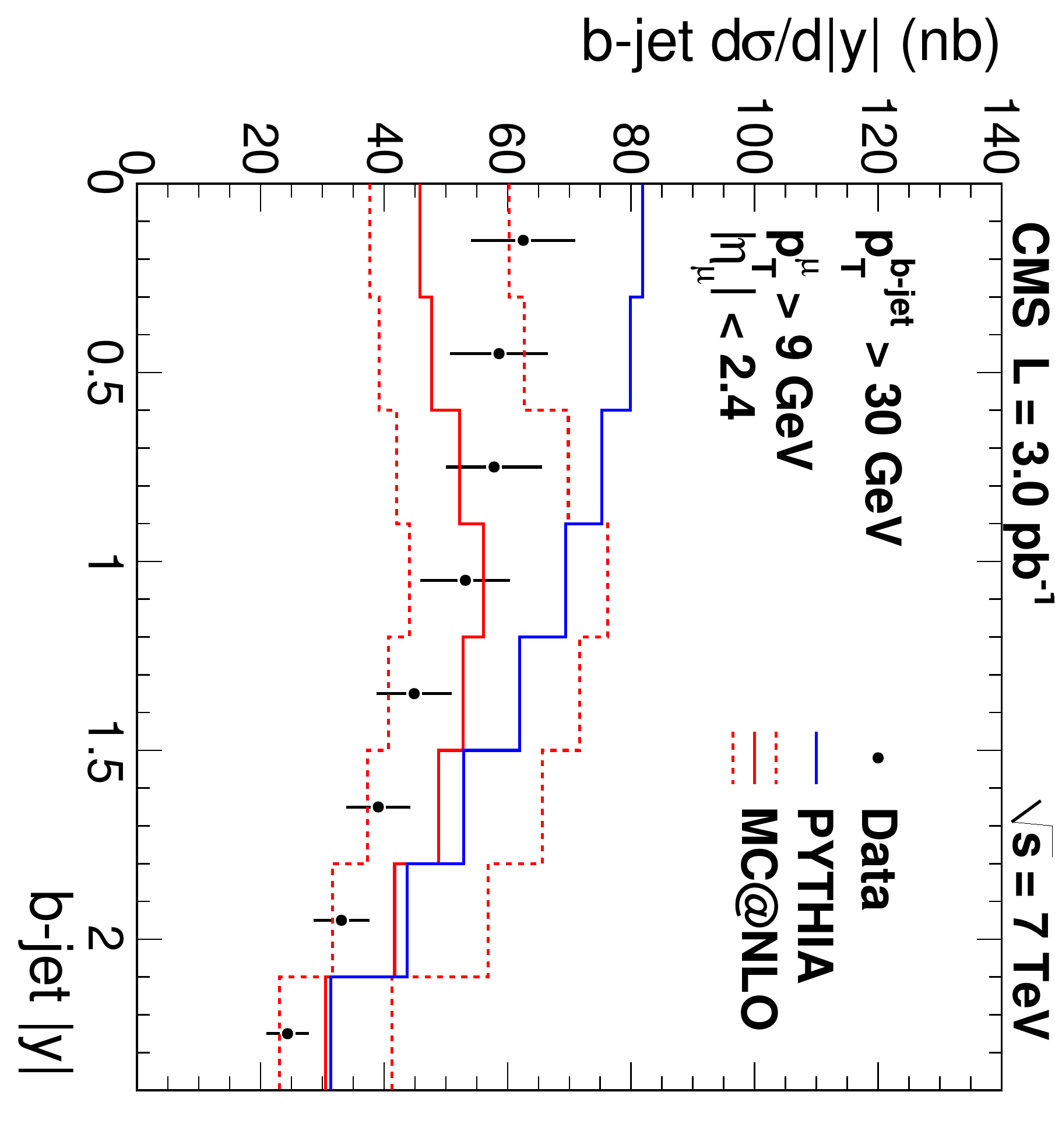}
\includegraphics[width=0.49\textwidth]{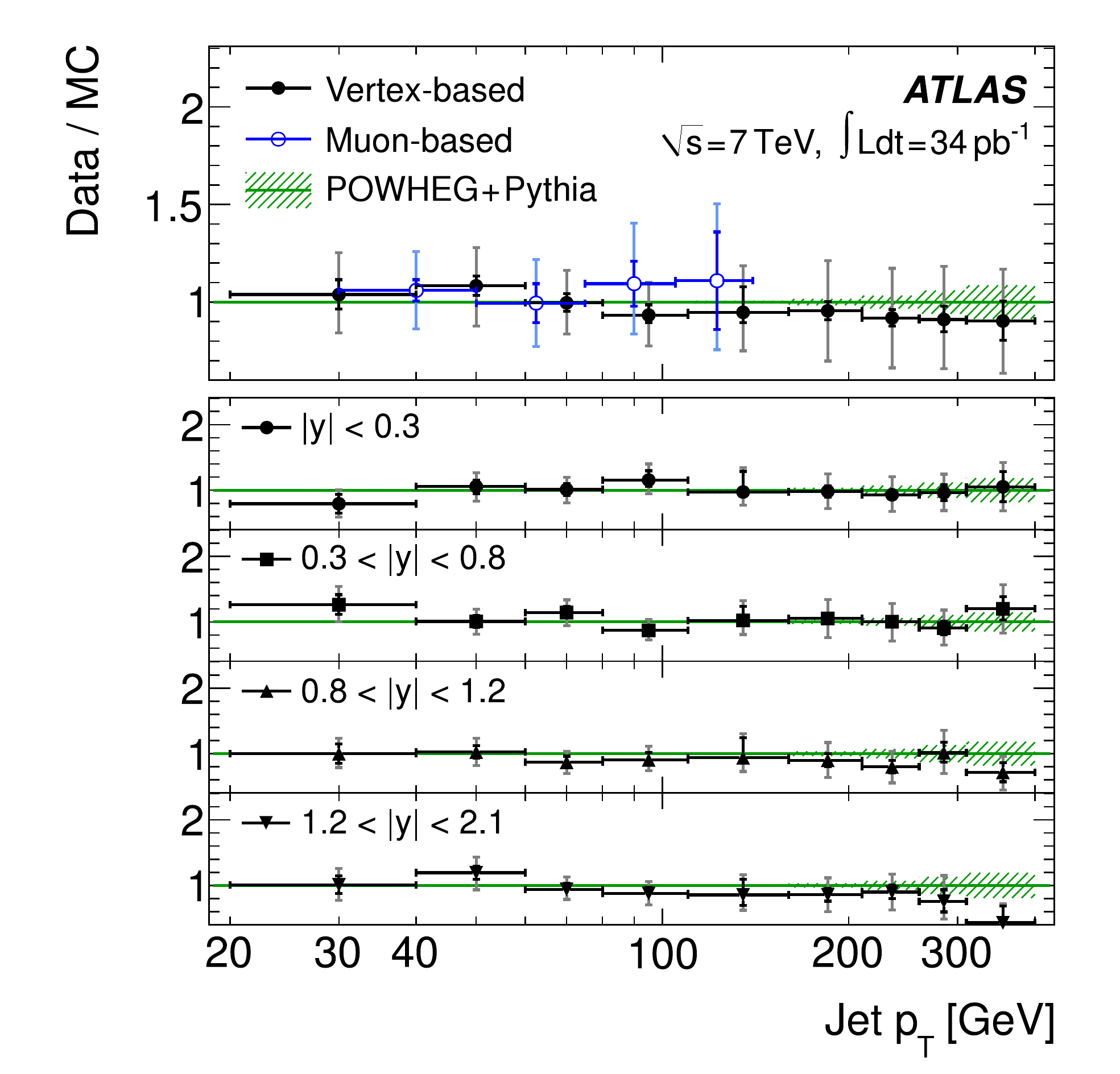}
\caption{\label{fig:b}
(Left) Differential b-jet cross section versus rapidity at CMS, showing comparisons to MC@NLO and Pythia.
\cite{Chatrchyan:2012dk}
(Right) Good agreement between data and Powheg+Pythia seen at ATLAS.
\cite{ATLAS:2011ac}
}
\end{figure}

The results for inclusive b-dijet production agree well with predictions from LO and NLO generators and NLO pQCD. The inclusive b-jet production shows some differences at high b-jet $p_T$, where gluon splitting becomes dominant. As shown in Fig.~\ref{fig:b}, the NLO generator Powheg interfaced with Pythia describes ATLAS data well, while MC@NLO has some disagreement on the differential distributions from CMS. The CMS and ATLAS data themselves agree well with each other. The total b-hadron and b-jet cross sections tend to lie between MC@NLO and Pythia predictions.

\begin{figure}[htbp!]
\centering
\includegraphics[width=0.35\textwidth]{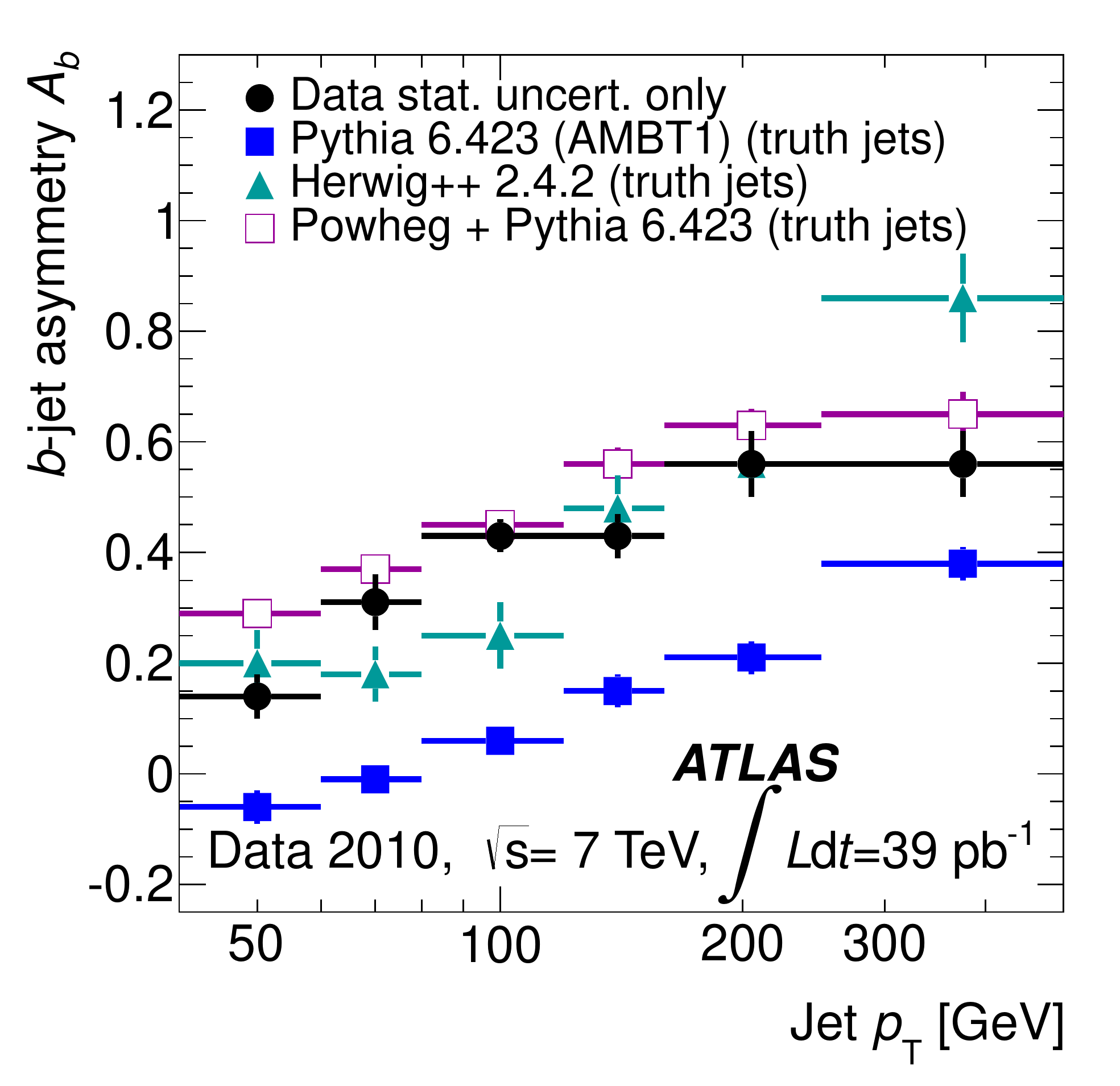}
\includegraphics[width=0.31\textwidth]{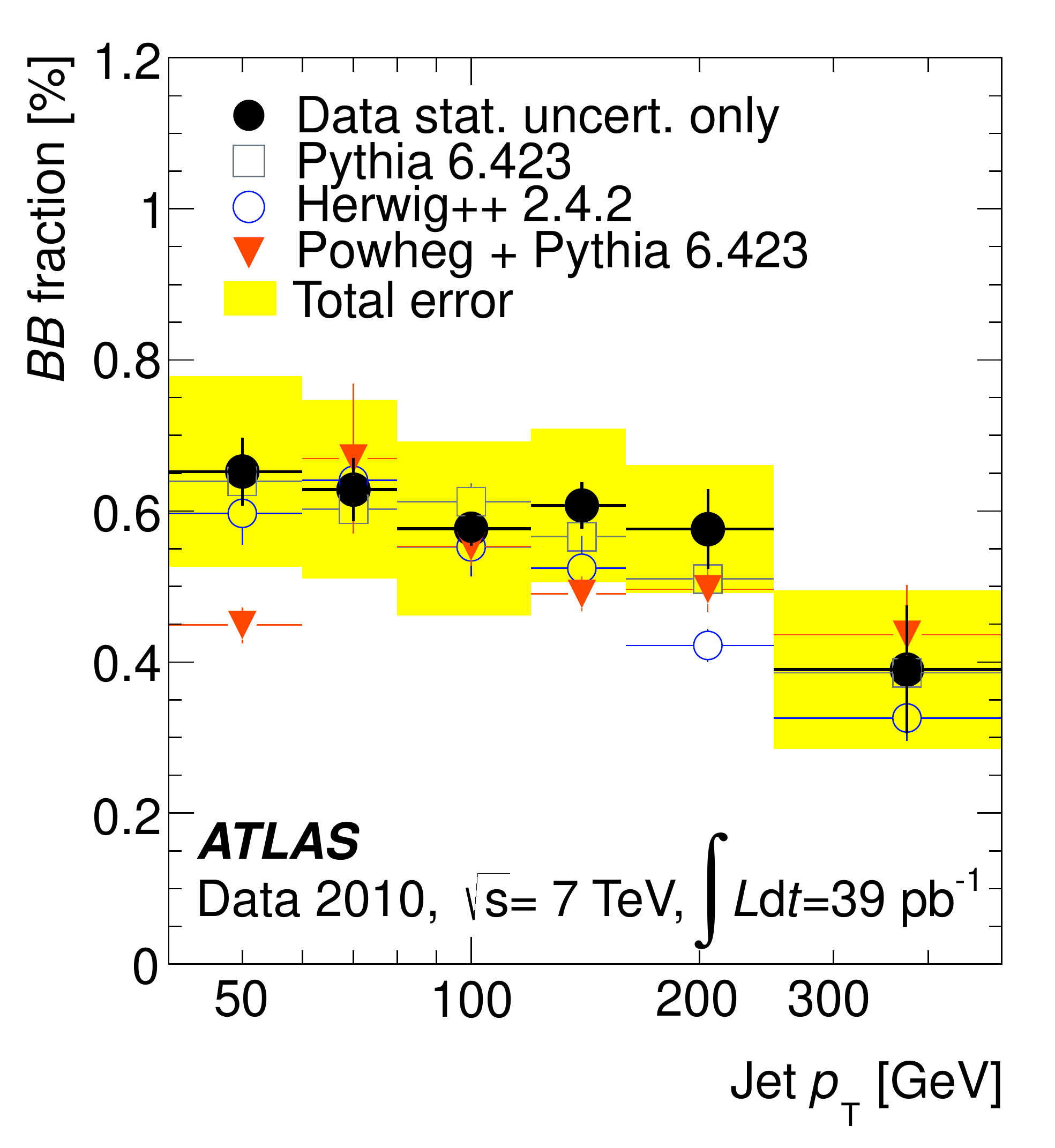}
\includegraphics[width=0.31\textwidth]{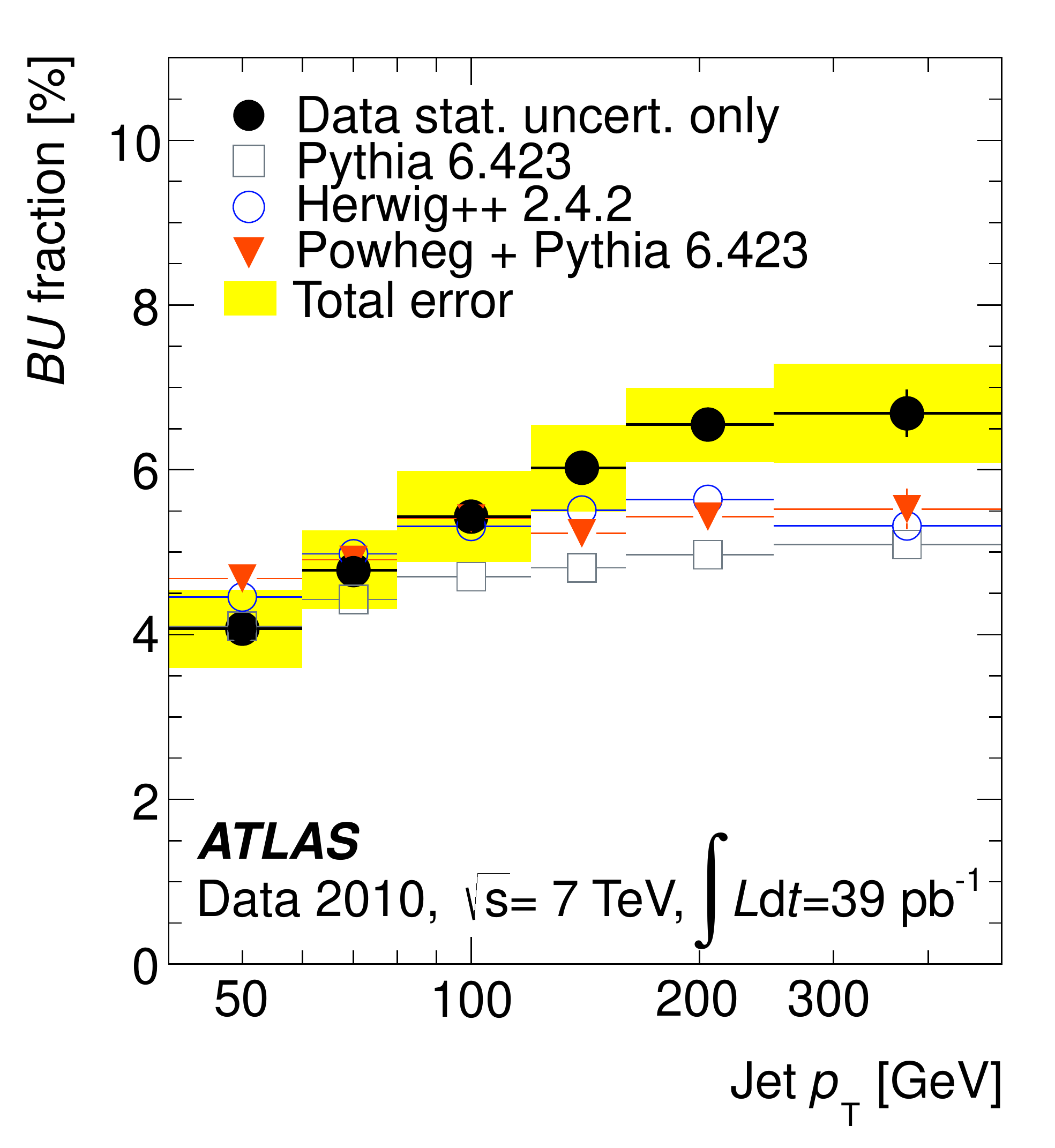}
\caption{\label{fig:bb}
(Left) The b-flavor asymmetry measured by ATLAS.
(Middle) Production fraction of dijets with two b jets, and
(right) with one b jet and one light jet.
\cite{Aad:2012ma}
}
\end{figure}

The dijet flavor composition measurement from ATLAS, shown in Fig.~\ref{fig:bb}, summarizes key weak points in the b-jet predictions. The asymmetry between b and non-b jets is defined as $A_b = N^{SL}_b/N^{L}_b-1$, where $N^{L}_b$ and $N^{SL}_b$ are the number of leading and subleading b jets, respectively. This asymmetry is quite large in data, which would also effect the inclusive b-jet spectrum. The disagreement with LO Pythia and agreement with NLO generator Powheg interfaced with Pythia suggests that NLO effects are important. The fraction of two b jets (BB) is well modeled as are all other fractions except b plus light flavor (BU), which shows disagreement at high jet $p_T$. The former is sensitive to flavor creation, while the latter is sensitive to flavor excitation and gluon splitting.
The underestimation of gluon splitting by theory is supported by the CMS study of b\bbar-hadron angular correlations \cite{Khachatryan:2011wq}.

\section{W+c}
\label{sec:Wc}

The W+c production at the LHC provides an important constraint for the strange quark PDF. Some analyses of the previous data from neutrino-nucleon deep inelastic scattering (DIS) experiments have suggested that the s-quark sea is suppressed relative to the d-quark sea at all values of Bjorken $x$ \cite{Martin:2009iq,Alekhin:2009ni,Ball:2008by} while others \cite{Lai:2010vv} suggest that the SU(3) symmetry is restored at small $x$. In addition, fits to the neutrino DIS data from NuTeV prefer a small unexpected asymmetry between the s and \sbar\ sea \cite{Goncharov:2001qe,Martin:2009iq,Ball:2009mk}.

These effects are best studied at the LHC with the W+c production, which has been measured by ATLAS \cite{Aad:2014xca} and CMS \cite{Chatrchyan:2013uja} using 7~TeV data.
Both measurements tag c with either a semileptonic decay (soft muon or electron in a jet) or a charmed meson (D$^{\pm}$ and D$^{*\pm}$). The basic analysis cuts for ATLAS (CMS) are W boson transverse mass $m_{T,\rm W}>40$~GeV, lepton $p_T^{l}>20$~(25--35)~GeV, $|\eta_{\rm lepton}|<2.5 (2.1)$, jet $p_T^{\rm jet}>25$~GeV and $|\eta_{\rm jet}|<2.5$. The main differences between the two analyses are in the lepton phase space. Comparisons to various PDFs are calculated with aMC@NLO (MCFM).

\begin{figure}[htbp!]
\centering
\includegraphics[width=0.25\textwidth]{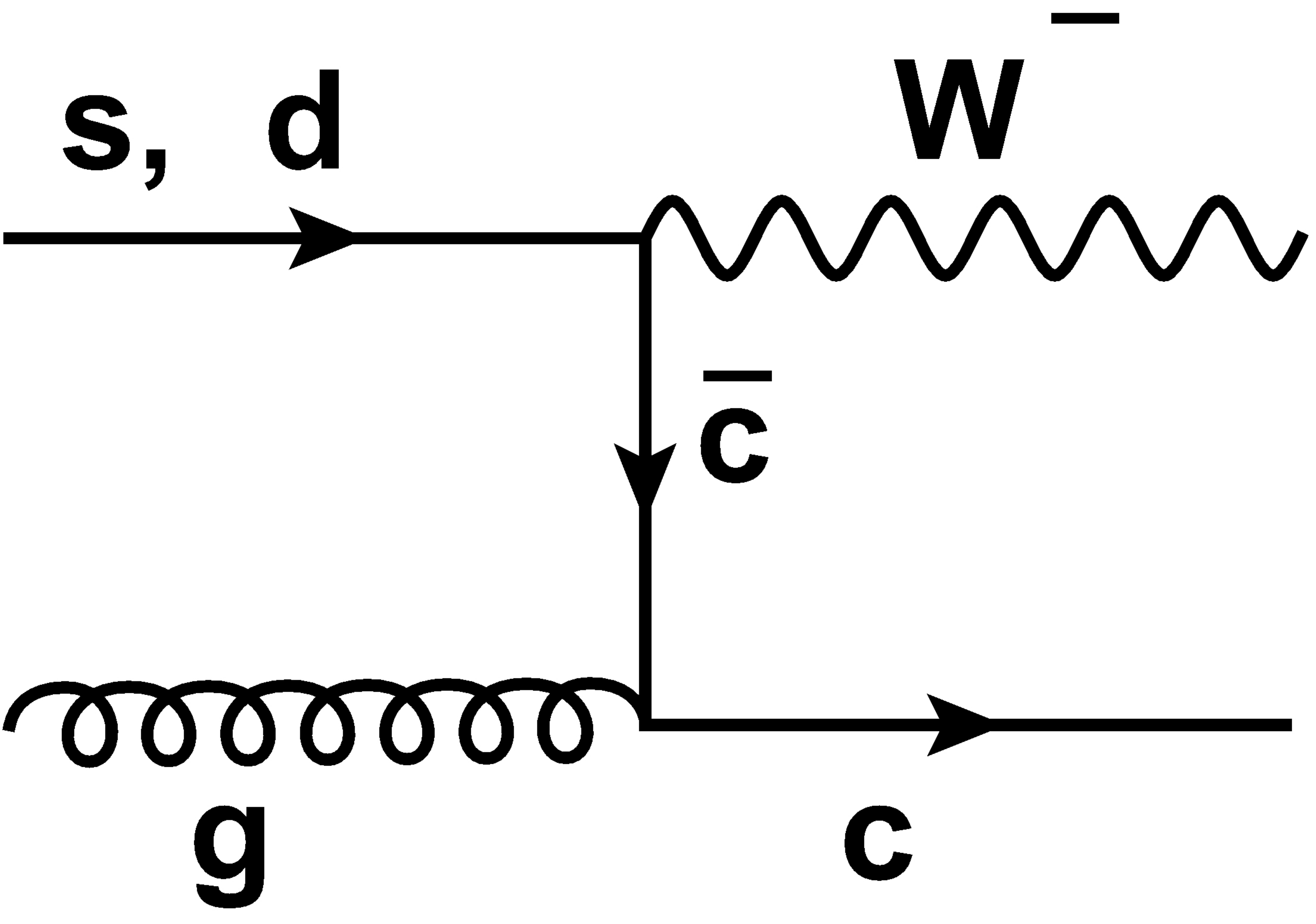}
\hspace{0.1\textwidth}
\includegraphics[width=0.25\textwidth]{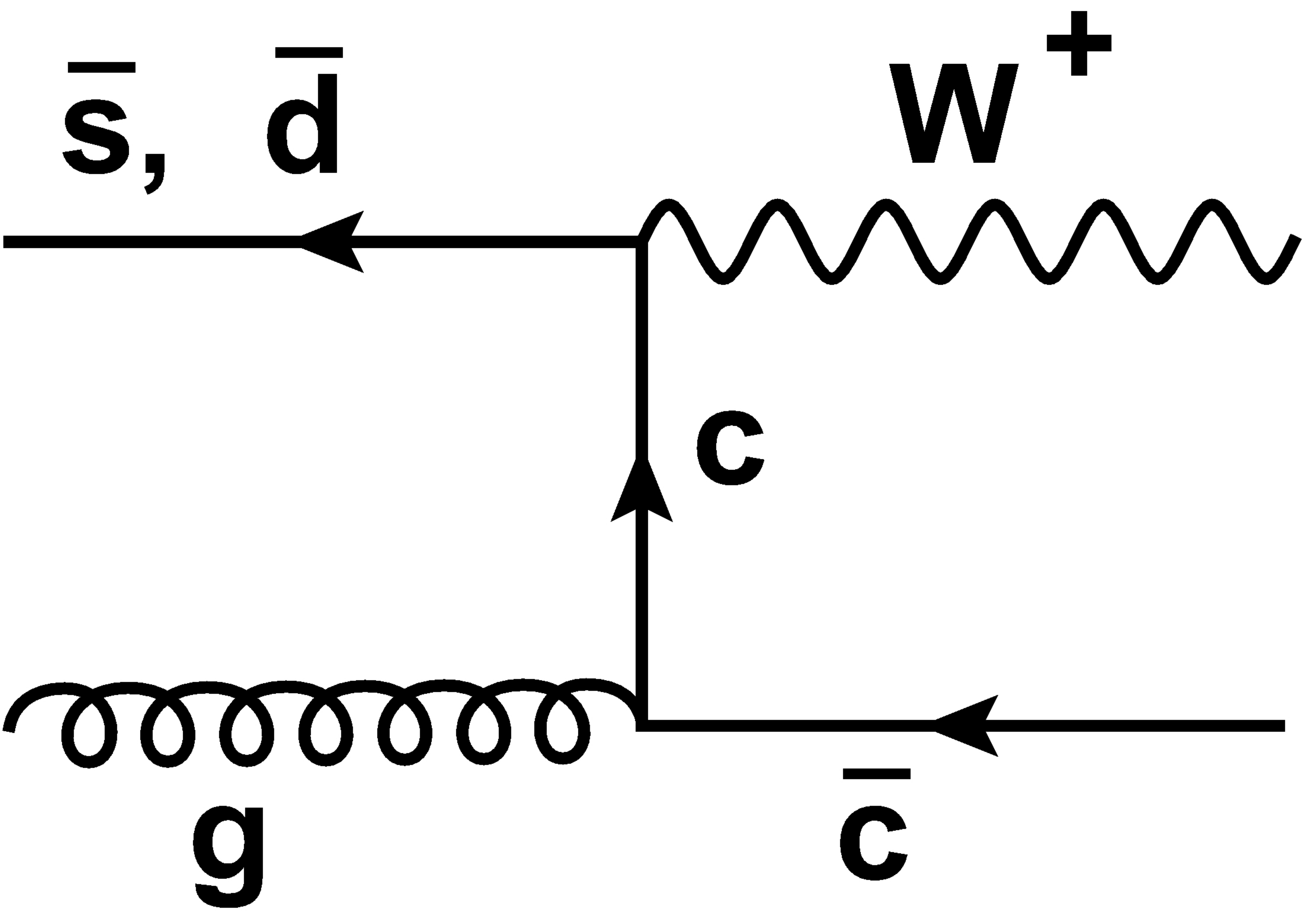}
\caption{\label{fig:wc}
Leading Feynman diagrams for W+c production. The d-channel is Cabibbo suppressed by $V_{\rm dc}$ and contributes about 10\%. Channels with initial state c and g$\rightarrow$c\cbar\ are suppressed by subtracting same sign background of W+c from the opposite sign signal region.
\cite{Chatrchyan:2013uja}
}
\end{figure}

The leading diagrams for W+c production shown in Fig.~\ref{fig:wc} are g\sbar$\rightarrow$W$^++$\cbar\ and gd$\rightarrow$W$^-+$\cbar\ and their charge conjugates. The gd process is Cabibbo suppressed by $V_{\rm dc}$ and only contributes about 10\% of the inclusive cross section. The key feature of W+c production is the opposite sign (OS) charges of the W and c, which allows suppressing charge-symmetric backgrounds by subtracting same sign (SS) background from the OS signal region. This also suppresses c jets from initial state charm and from g$\rightarrow$c\cbar, making W+c very sensitive to the strange quark PDF.
A small asymmetry of 5--10\% between W$^++$\cbar\ and W$^-+$c production is expected to arise from the valence-d contributions in the gd channel, increasing at high $|\eta|$ with increasing $x$. Any residual asymmetry relative to pQCD predictions with s$=$\sbar\ sea could be interpreted as s$-$\sbar\ asymmetry.

\begin{figure}[htbp!]
\centering
\includegraphics[width=0.47\textwidth]{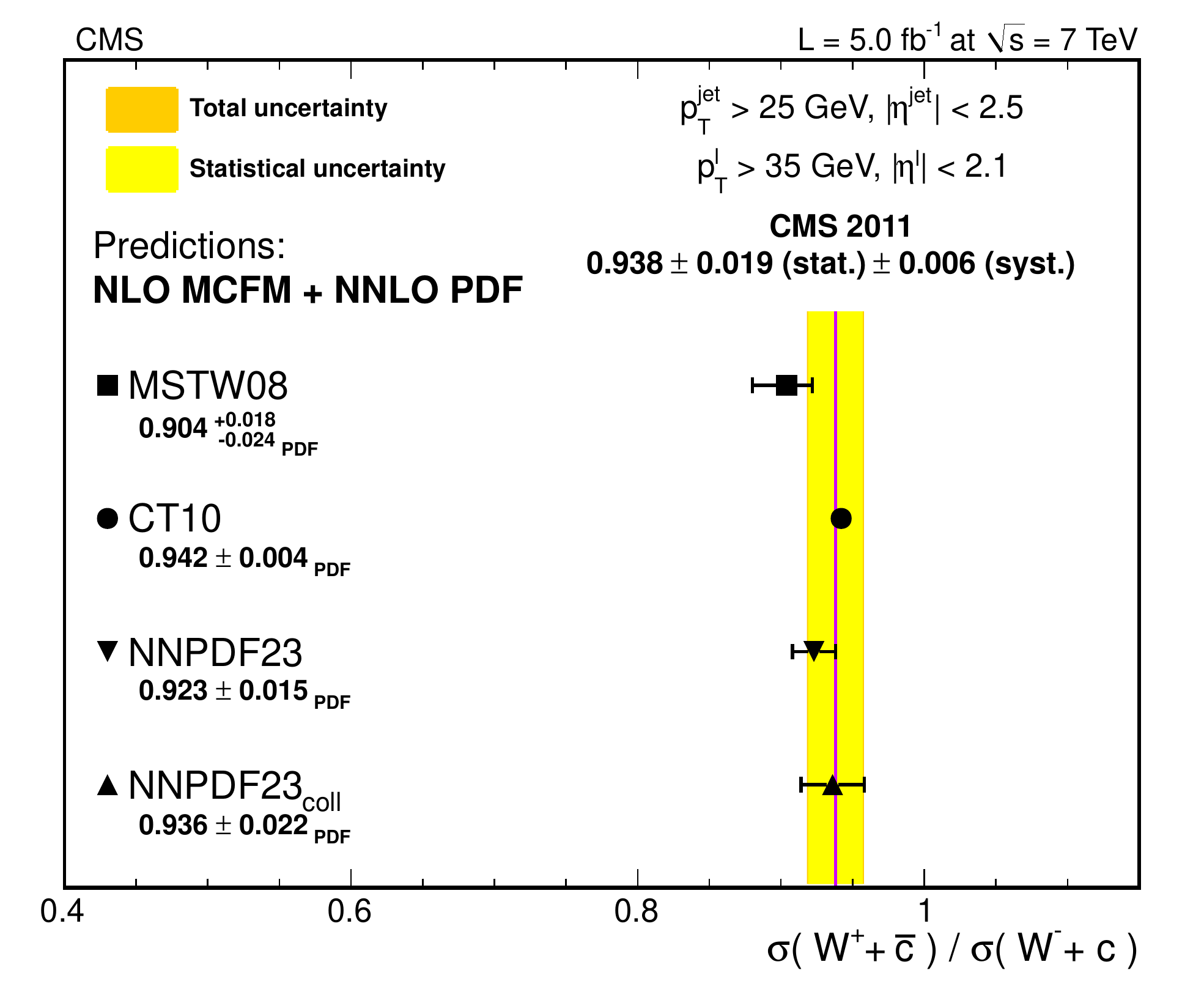}
\includegraphics[width=0.52\textwidth]{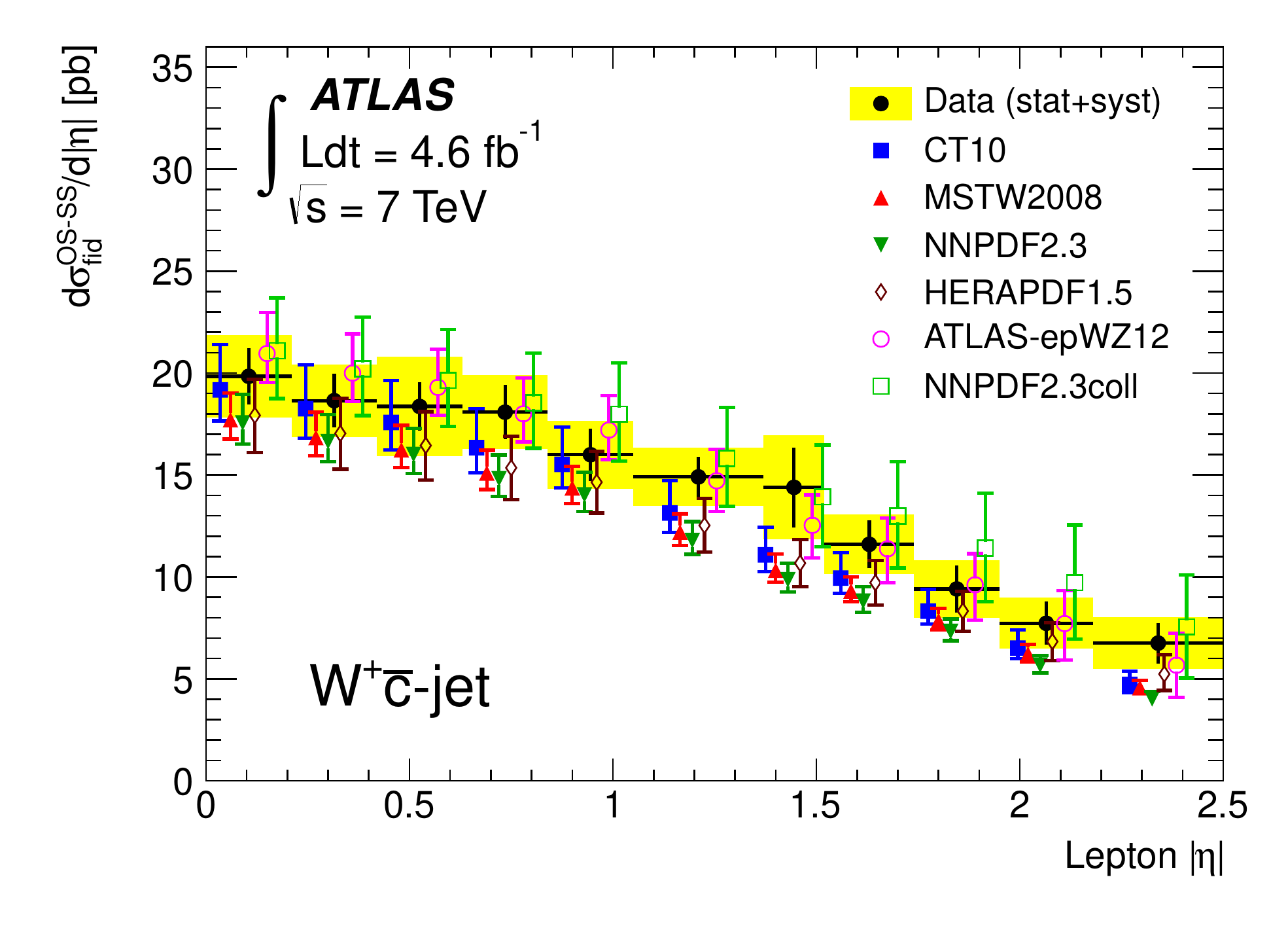}
\caption{\label{fig:wcxsec}
(Left) Ratio
of $\sigma({\rm W}^-+{\rm c})/\sigma({\rm W}^++\cbar)$ cross sections at CMS, confirming the expected asymmetry arising from d valence quarks in the gd channel,
\cite{Chatrchyan:2013uja} and
(right) measurement of the fiducial W$^-+$c cross section at ATLAS.
\cite{Aad:2014xca}
}
\end{figure}

Both experiments confirm the expected $\sigma({\rm W}^++\cbar)/\sigma({\rm W}^-+{\rm c})$ asymmetry from d-valence at the 2--3~$\sigma$ level, as shown in Fig.~\ref{fig:wcxsec}(left), and also demonstrate sensitivity to the strangeness PDF in global fits that also include W and Z measurements \cite{Aad:2012sb,Chatrchyan:2013mza} for further u and d valence quark constraints. Neither experiment is yet sensitive to the expected s$-$\sbar\ differences implemented in various PDFs, but ATLAS quotes $A_{\rm s\sbar}=(2\pm 3)$\% relative to CT10 that implements a s$=$\sbar\ sea, with the quoted uncertainty dominated by statistical uncertainties and comparable to the s$-$\sbar\ differences implemented in the MSTW2008 PDF.

The W+c cross sections measured by CMS and ATLAS disagree by about 2~$\sigma$ relative to each other: ATLAS' fiducial W+c cross section is 10--15\% higher than the CT10 prediction and most PDFs except ATLAS-epWZ12 and NNPDF2.3COLL, as shown {\em e.g.} in Fig.~\ref{fig:wcxsec}(right), while CMS' W+c cross section
is consistent with CT10,
but disagrees with NNPDF2.3COLL.
The conclusions drawn by the two experiments consequently differ somewhat: ATLAS supports the hypothesis of a SU(3) symmetric light quark sea with $r_{\rm s}=0.96$, while CMS is consistent with the lower value of $r_{\rm s}$ obtained by the NOMAD experiment \cite{Samoylov:2013xoa} and used by most mainstream PDFs.

\section{V+b}
\label{sec:Vb}

The W+b(b) and Z+b(b) channels, jointly referred to as V+b, provide an important test of pQCD in the presence of heavy quarks. The gauge boson and b masses probe different scales, and the V+b offers a chance to test different flavor and mass schemes. The production of heavy gauge bosons in association with b quarks also constitutes an irreducible background for the search for the Higgs decays to H$\rightarrow$b\bbar\ and for various BSM processes.

Due to the suppression of direct Wbq vertices for energy regimes below the top quark mass, W+b(b) and Z+b(b) signatures probe slightly different physics. Whereas W+b\bbar\ production originates mostly from final-state gluon splitting, signatures with a Z and a non-collinear b\bbar\ pair are also produced from gluon initial states. The production of a Z boson with a single b quark therefore probes the b-quark content of the proton. Both W and Z can also be produced from a bq initial state in association with a b and a light-flavored quark.

First measurements of the Z+b and W+b cross sections at the Tevatron found some tension between the data and NLO calculations or predictions from ME+PS generators \cite{Aaltonen:2009qi,D0:2012qt,Aaltonen:2008mt,Abazov:2013uza}.
The new measurements from the LHC are therefore important to resolve these disagreements with the theory predictions.

\section{W+b production}
\label{sec:Wb}

The large t\tbar\ and single t backgrounds are the main experimental challenge for the W+b measurements, which explicitly exclude contributions from t$\rightarrow$Wb.
The W+b measurements at ATLAS \cite{Aad:2011kp,Aad:2013vka} and CMS \cite{Chatrchyan:2013uza} use complementary approaches: ATLAS measures W+b from events with exactly one b-tagged jet and at most two jets to avoid the large t\tbar\ background, while CMS selects exactly two well-separated b-tagged jets to avoid the theoretically difficult region of collinear b\bbar. The ATLAS measurement is later unfolded to exclusive W+b(1jet) and W+b(2jet) final states, the latter of which includes the W+2b final state.

The leading production mode of W+b in the standard model is gluon-splitting in q'\qbar$\rightarrow$W(g)b\bbar\ where the b quarks can be collinear and reconstructed as a single jet. Significant contributions to the W+b production also come from double parton scattering (DPS), where the W and b\bbar\ come from different parton interactions. Both experiments compare their measurements to NLO pQCD predictions with corrections for non-perturbative effects (hadronization and underlying event) and DPS.

\begin{figure}[htbp!]
\centering
\includegraphics[width=0.54\textwidth]{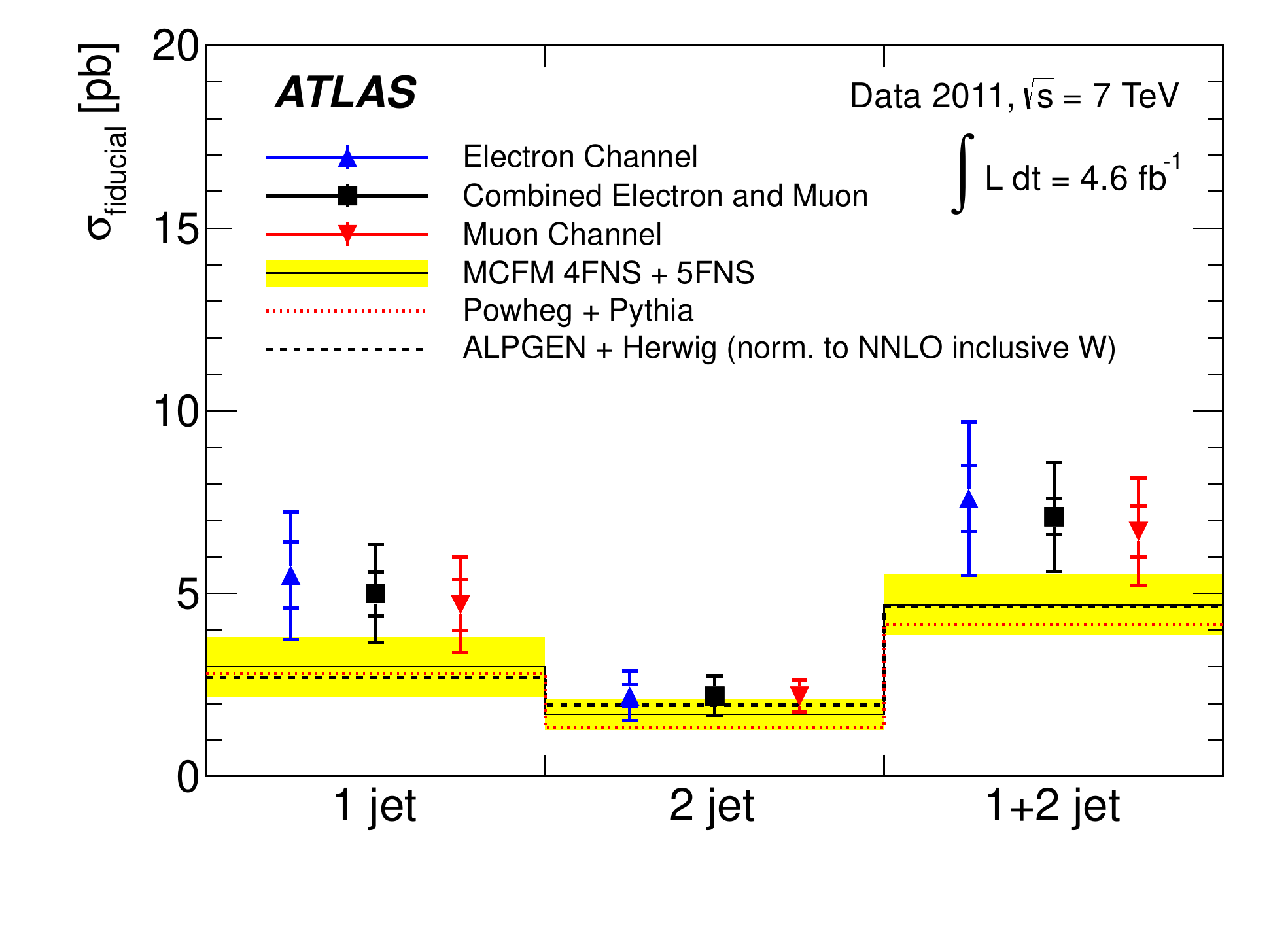}
\includegraphics[width=0.45\textwidth]{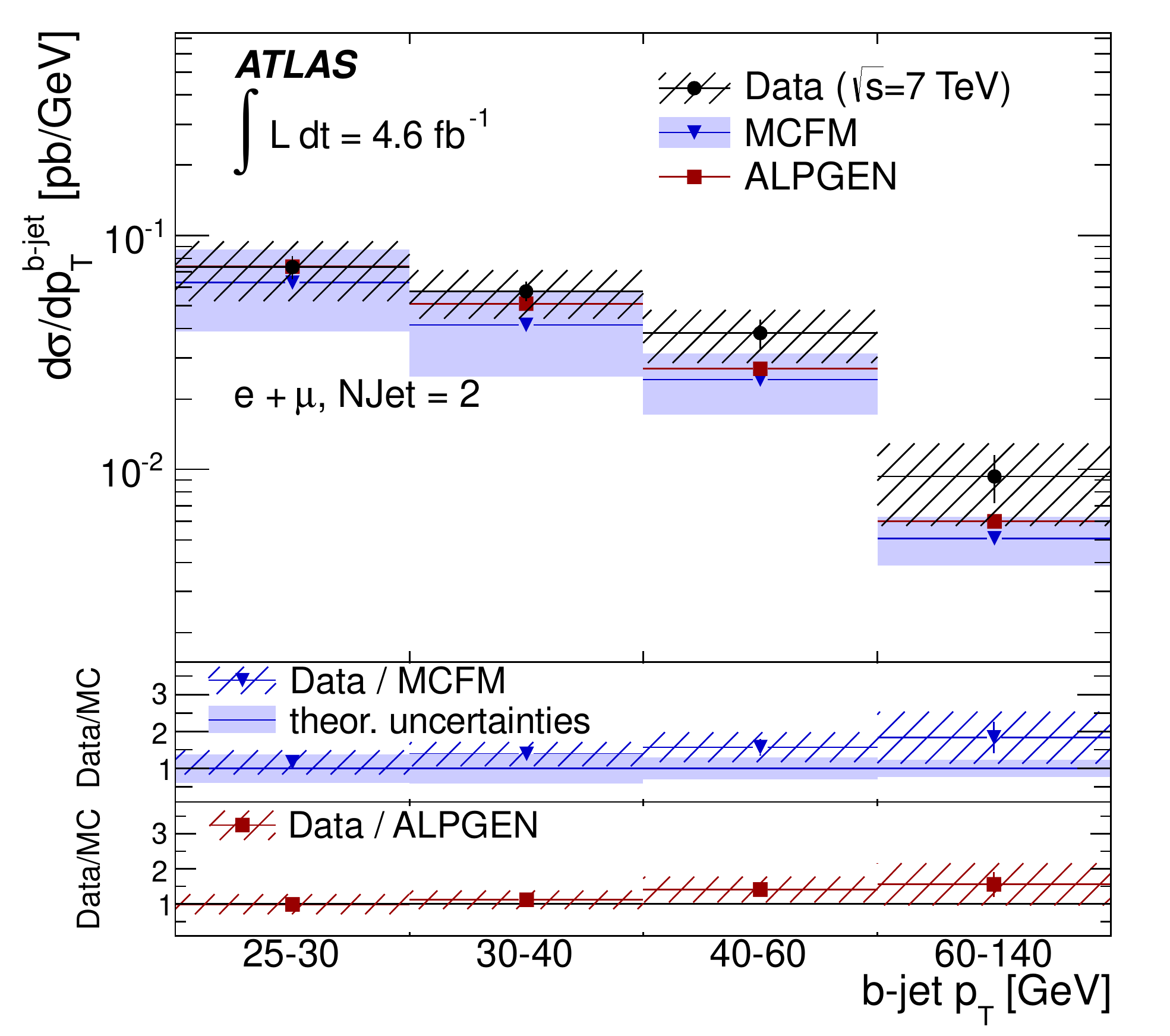}
\caption{\label{fig:wb}
Measurement of the fiducial W+b cross section at ATLAS versus number of jets (left) and versus b-jet $p_T$ for W+b(2jet) (right)
\cite{Aad:2013vka}
}
\end{figure}

CMS finds agreement between data and theory for the W+2b production, as does ATLAS for the W+b(2jet) production, which is very sensitive to W+2b.
The ATLAS measurement also looks at b-jet $p_T$ differentially,
as shown in Fig.~\ref{fig:wb},
while CMS investigates multiple kinematic distributions with W+b backgrounds included. The W+b(1jet) channel studied at ATLAS shows a higher cross section than the theory predictions, with agreement worsening at high b-jet $p_T$ in both W+b(1jet) and W+b(2jet) channels.
The overall data/theory disagreement on the total fiducial cross section is about $1.5~\sigma$, dominated by the W+1b channel.
There is consistency between the trend of too low cross sections modeled for W+b in the ATLAS measurement and the trend to underestimate the rate of collinear gluon splitting in simulations. The collinear b\bbar\ is often reconstructed as one jet in W+b.

ATLAS has also made a second set of measurements with the combined cross section of W+b plus single top quarks, and finds significantly reduced uncertainty for the b-jet $p_T$ differential cross sections with respect to single-top subtracted measurements. They can be compared to combined single top-quark and W+b calculations in the future.

\section{Z+b production}
\label{sec:Zb}

The Z+b production has been measured
in 7~TeV pp collisions by CMS \cite{Chatrchyan:2012vr,Chatrchyan:2013zja,Chatrchyan:2014dha} and ATLAS \cite{Aad:2011jn,Aad:2014dvb}. The first pp measurements \cite{Chatrchyan:2012vr,Aad:2011jn} focused on the total Z+b-jet cross section,
while later papers \cite{Chatrchyan:2014dha,Aad:2014dvb} have looked at separate Z+1 b-jet and Z+2 b-jet production. One study \cite{Chatrchyan:2013zja} looked at angular distributions of b hadrons instead of b jets, allowing for a better study of the quasi-collinear region.

\begin{figure}[htbp!]
\centering
\includegraphics[width=0.32\textwidth]{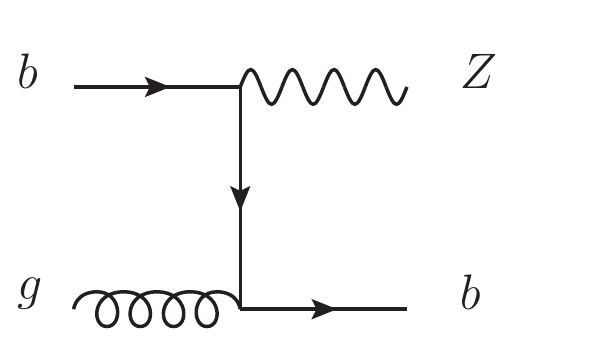}
\includegraphics[width=0.32\textwidth]{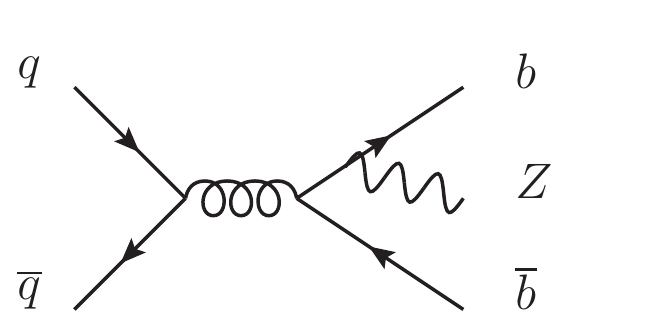}
\includegraphics[width=0.32\textwidth]{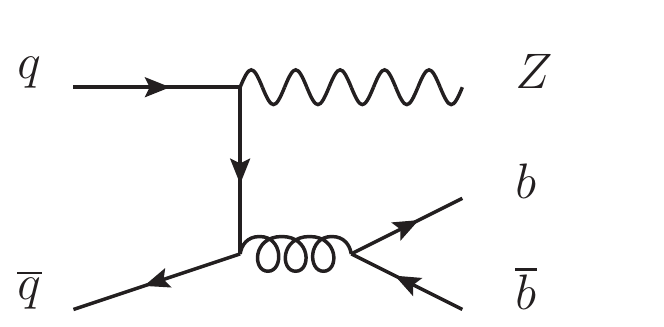}
\caption{\label{fig:zb}
Leading Feynman diagrams for Z+b production. The first diagram is only available in 5 FNS, the other two in both 5 FNS and 4 FNS.
\cite{Aad:2014dvb}
}
\end{figure}

The b-jet production in association with a Z boson can be modeled in either the 4FNS (also known as fixed flavor scheme), or in the 5FNS (also known as variable flavor scheme). In the 4FNS the proton PDFs only contain the four lightest quark flavors (u, d, s, c) and the b quarks are produced from an explicit gluon splitting g$\rightarrow$b\bbar. The 4FNS calculations usually include the b quark mass. The 5FNS considers b quarks as the fifth flavor in the proton PDFs, which effectively allows singly-produced b quarks. The available 5FNS calculations do not yet consider b quark masses. Both schemes agree in an all orders calculation, but can differ at finite order of pQCD. The leading Feynman diagrams of Z+b-jet production are shown in Fig.~\ref{fig:zb}.

\begin{figure}[htbp!]
\centering
\includegraphics[width=0.49\textwidth]{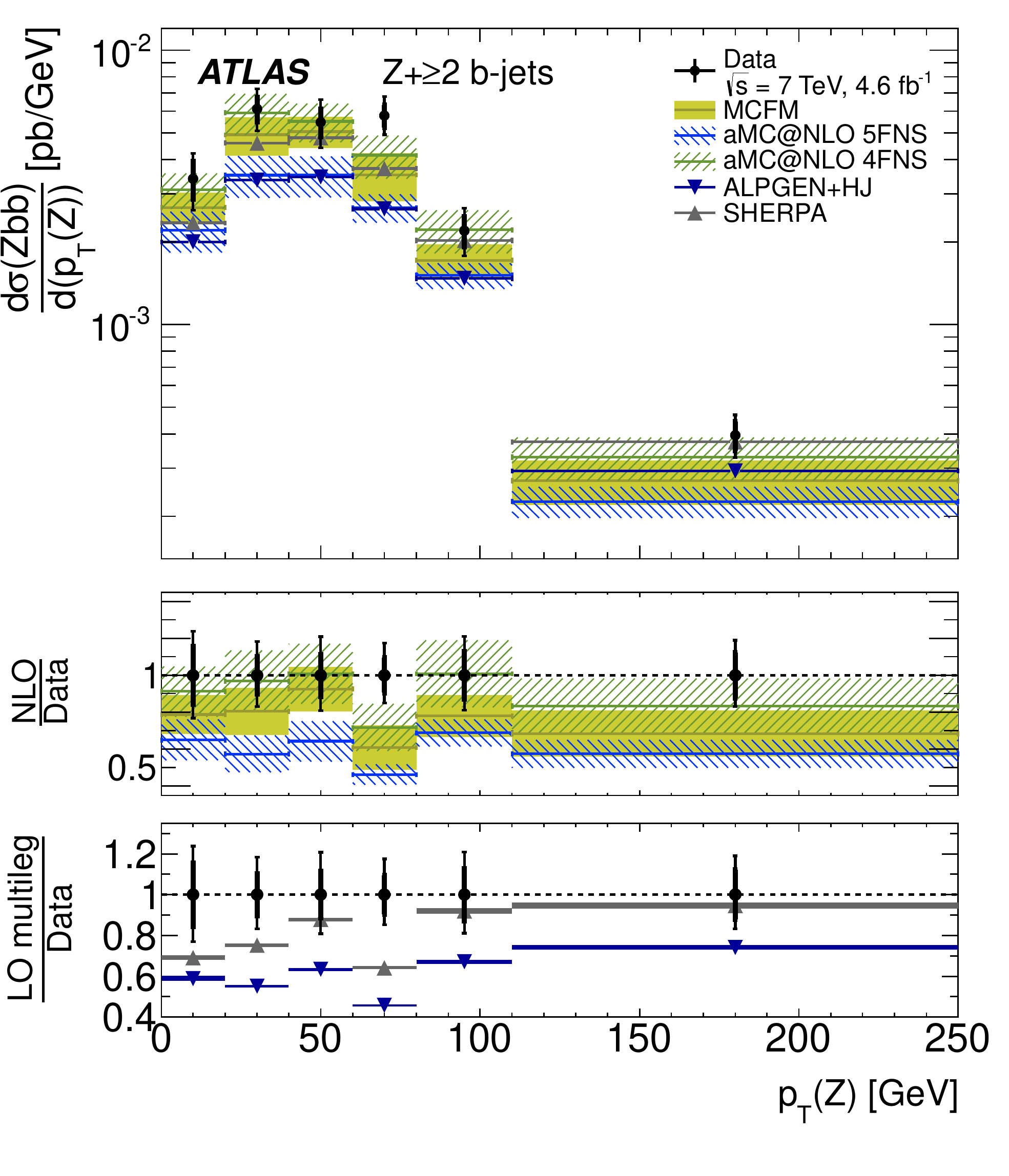}
\includegraphics[width=0.49\textwidth]{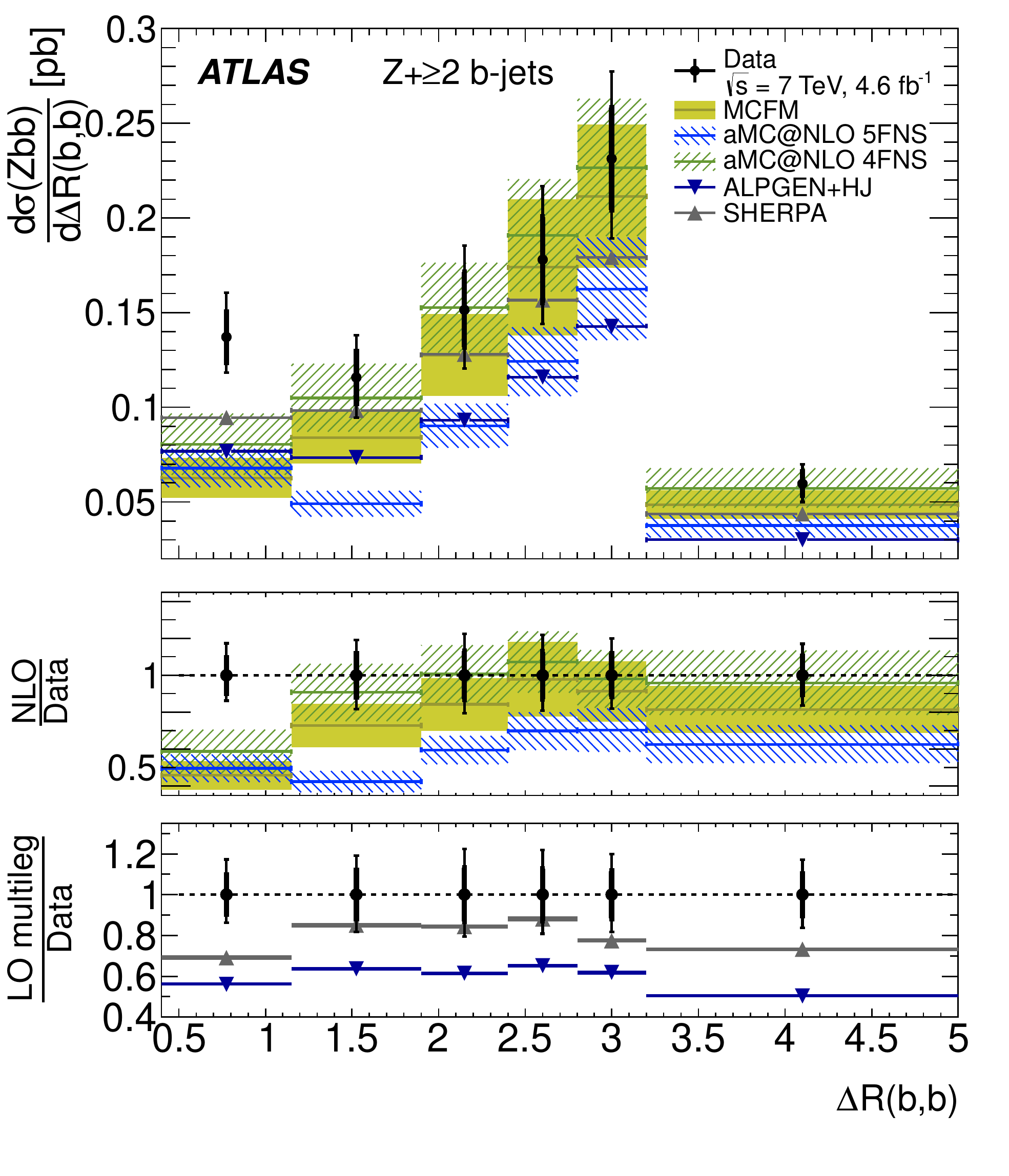}
\caption{\label{fig:zpt}
(Left) Measurement of Z boson $p_T$ for Z+$\geq2$ b jets.
(Right) Angular distribution of the leading b jets in Z+$\geq2$ b-jet events.
\cite{Aad:2014dvb}
}
\end{figure}

Particular focus in the Z+b-jet studies has been given to the $p_T$ of the Z boson and b\bbar\ systems, $p_{T}({\rm Z})$ and $p_{T}({\rm b\bbar})$, as well as to the angular separation $\Delta R({\rm b},\bbar)$ and mass $m({\rm b},\bbar)$ of the di-b-jet system, all of which are important for the Z(H$\rightarrow$b\bbar) channel.
 The $p_{T}({\rm Z})$ for Z+$\geq2$b-jet and $\Delta R({\rm b},\bbar)$ are shown in Fig.~\ref{fig:zpt}.
The general findings from the recent studies are that 4FNS calculations predict too few Z+1 b-jet events, but are better than 5FNS for Z+2 b-jet events. The 5 FNS is overall fine, but tends to produce a too soft $p_{T}({\rm Z})$ spectrum in Z+$\geq2$b-jet events, presumably due to missing quark-mass effects. The multileg LO generators do quite well on the shape of the angular distributions, but require additional correction factors to match NLO production. The region of collinear b jets (small $\Delta R({\rm b},\bbar)$) tends to be particularly difficult for all approaches, although AlpGen in 4FNS performs somewhat better than other generators.

\section{Boosted Z$\rightarrow$b\bbar}
\label{sec:Ztobb}

Most Higgs bosons produced at the LHC decay into a pair of b jets, but finding them from the overwhelming QCD background of b jets is very difficult.
Selecting pairs of boosted ({\em e.g.} $p_{T}({\rm b},\bbar)>200$~GeV) b jets with associated forward jets may allow the discovery of H$\rightarrow$b\bbar\ events in the vector boson fusion (VBF) mode. The boosted b\bbar\ topology is also used in searches of associated standard model Higgs boson production in the VH(b\bbar{}) and t\tbar{}H(b\bbar{}) modes, as well as for searches of TeV-scale resonances decaying to b\bbar b\bbar\ via ZZ, ZH or HH.
Finding fully hadronic Z$\rightarrow$b\bbar\ is the first important step on the way. The boosted Z$\rightarrow$b\bbar\ can also provide a useful benchmark {\em e.g.} for b-JES, as demonstrated at the Tevatron~\cite{Donini:2008nt}.

\begin{figure}[htbp!]
\centering
\includegraphics[width=0.49\textwidth]{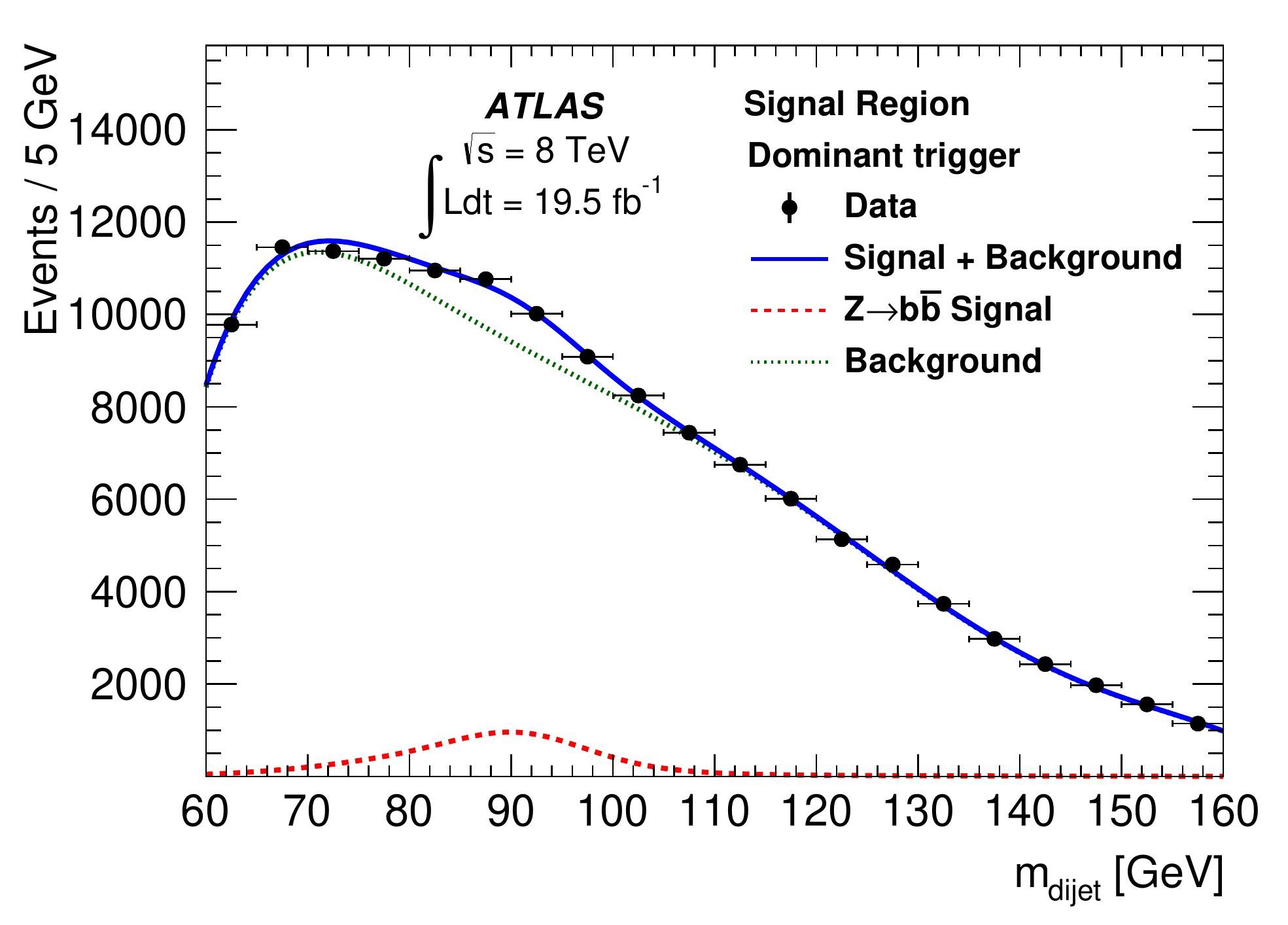}
\includegraphics[width=0.49\textwidth]{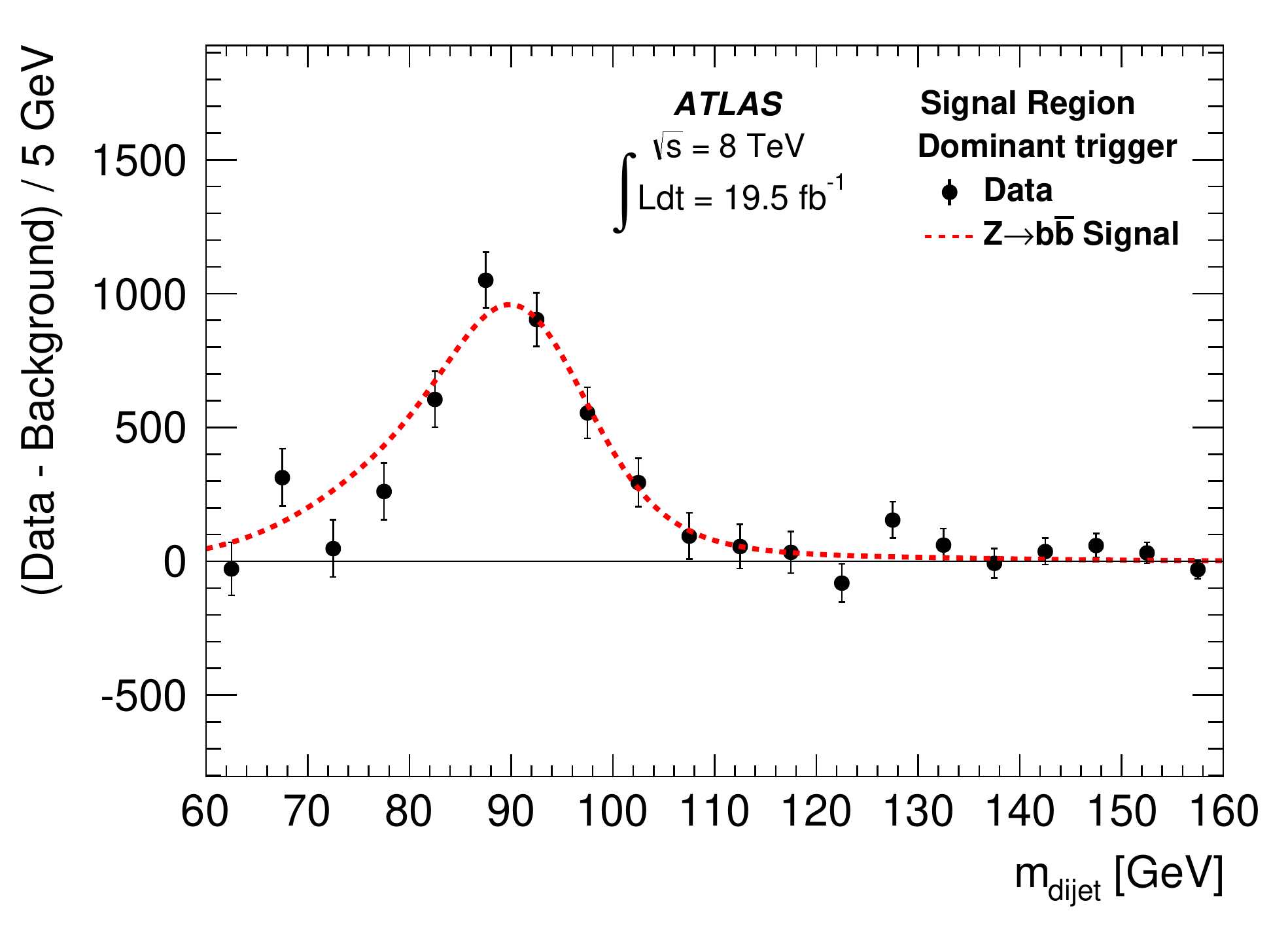}
\caption{\label{fig:ztobb}
Extraction of Z$\rightarrow$b\bbar\ from the signal-enriched b-tagged dijet mass distribution (left), with a clear signal peak seen for boosted Z$\rightarrow$b\bbar\ (right).
\cite{Aad:2014bla}
}
\end{figure}

The cross section of high transverse momentum Z$\rightarrow$b\bbar\ production at the LHC has been first measured by ATLAS \cite{Aad:2014bla}. Figure~\ref{fig:ztobb} shows the result of a simultaneous fit to signal (illustrated) and control regions, with Z$\rightarrow$b\bbar\ evident as a clear peak over the background model. The dominant background is multijet events, with small contributions from Z$\rightarrow$c\cbar, t\tbar\ and W$\rightarrow$q\qbar'. The measured fiducial cross section $\sigma^{\rm fid}_{{\rm Z}\rightarrow{\rm b\bbar}}=2.02\pm 0.33$~pb is in good agreement with NLO-plus-parton-shower predictions from Powheg+Pythia of $\sigma^{\rm fid}_{{\rm Z}\rightarrow{\rm b\bbar}}=2.02\pm 0.25$~pb and aMC@NLO+Herwig++ of $\sigma^{\rm fid}_{{\rm Z}\rightarrow{\rm b\bbar}}=1.98\pm 0.16$~pb.
It is further found that the signal peak is consistent with the Z$\rightarrow$b\bbar\ expectation: $\delta M_{\rm Z} = -1.5 \pm 3.5$~GeV.

\section{Conclusions and outlook}
\label{sec:summary}

The heavy flavor jet measurements presented in this article are summarized in Table~\ref{tab:summary}, which indicates the integrated luminosity and $\sqrt{s}$ used in these measurements by CMS and ATLAS.

\begin{table}[hbtp!]
\centering
\tbl{Summary of heavy flavor jet measurements at CMS and ATLAS reported in this review. The data sets are referred to as A (2010, 40~pb$^{-1}$, 7~TeV), B (2011, 5~fb$^{-1}$, 7~TeV), and C (2012, 20~fb$^{-1}$, 8~TeV). In some cases only a fraction of the full luminosity has been used.}
{\begin{tabular}{@{}cccccccc@{}}
\toprule
 & jet+b & W+c & W+b & Z+b & Z$\rightarrow$b\bbar & b-tag & b-JES \\
\colrule
CMS &
A\cite{Chatrchyan:2012dk} &
B\cite{Chatrchyan:2013uja} &
B\cite{Chatrchyan:2013uza} &
B\cite{Chatrchyan:2012vr,Chatrchyan:2014dha,Chatrchyan:2013zja} &
- &
{B\cite{Chatrchyan:2012jua}}, C\cite{CMS-PAS-BTV-13-001} &
{A\cite{Chatrchyan:2011ds}}, {B\cite{CMS-DP-2012-006}}, C\cite{CMS-PAS-JME-13-001} \\
ATLAS &
{A\cite{ATLAS:2011ac}}, B\cite{Aad:2012ma} &
B\cite{Aad:2014xca} &
{A\cite{Aad:2011kp}}, B\cite{Aad:2013vka} &
{A\cite{Aad:2011jn}}, B\cite{Aad:2014dvb} &
C\cite{Aad:2014bla} & 
{A\cite{ATLAS-CONF-2010-042,ATLAS-CONF-2010-091,ATLAS-CONF-2011-089}}, {B\cite{ATLAS-CONF-2011-102,ATLAS-CONF-2012-043,ATLAS-CONF-2012-097}}, C\cite{ATLAS-CONF-2014-004} & 
B\cite{Aad:2014bia} \\ 
\botrule
\end{tabular} \label{tab:summary}}
\end{table}

The b-tagging efficiencies and light-parton misidentification probabilities are found to be quite well modeled by simulation for various taggers, with scale factors within 10\% and 25\% from unity, respectively, and uncertainties 2--4\% and 8--17\%. The high performance taggers have efficiencies of the order 70\% with misidentification probability around 1\% for a medium working point, making light-parton backgrounds a small contribution for most analyses.

The b-JES is found to be very similar to inclusive JES, when excluding neutrinos from particle jets. Neutrinos carry on average about 5\% of the b-quark $p_T$ in inclusive b jets, and up to 15\% in semileptonic decays tagged with muons. The uncertainties on b-JES based on MC generator comparisons are 0.5--1.5\%, and they have been validated to about 0.5\% precision relative to inclusive jets with Z+b events.

Measurements of c jets in association with W bosons have proven to provide a powerful constraint on the proton strange quark sea at the LHC, but some discrepancies between experiments at the LHC and outside LHC still remain to be solved. The hypothesis of a $SU(3)$ symmetric light quark sea at low $x$ and an asymmetry between s and \sbar\ still remain to be conclusively demonstrated.

Studies of b-jet production with or without associated W and Z bosons find the di-b-jet $p_T$ and mass spectra to be well modeled by most generators on the market. However, sizable differences between data and predictions are seen in the modeling of events with single b jets, particularly at large b-jet $p_T$, where gluon splitting processes become dominant. This same source is also confirmed by studies of b-hadron and b-jet angular correlations as well as dijet flavor asymmetry and composition. None of the available theory predictions provides a fully satisfactory description of data with single b jets, but Powheg+Pythia seems most promising.

The first heavy flavor jet measurements with 8~TeV data have demonstrated the ability to measure boosted decays of Z$\rightarrow$b\bbar, paving the way for searches of H$\rightarrow$b\bbar\ and exotic new particles decaying into b jets.

\bibliographystyle{ws-ijmpa}
\bibliography{HeavyQuarkJets}

\end{document}